\documentclass[12pt]{article}
\usepackage{amsmath}
\usepackage{amsfonts}
\usepackage{geometry}
\usepackage{xcolor}
\usepackage{bbm}
\usepackage[onehalfspacing]{setspace}
\usepackage{graphicx}
\usepackage{url}
\usepackage{float}
\usepackage{natbib}
\usepackage{comment}
\newtheorem{theorem}{Theorem}

\newtheorem{proposition}[theorem]{Proposition}

\newenvironment{proof}[1][Proof]{\noindent\textbf{#1.} }{\ \rule{0.5em}{0.5em}}
\newcommand{\E}{\mathbb{E}}

\begin{document}
\allowdisplaybreaks

\title{Robust Econometrics for Growth-at-Risk\footnote{We benefited from valuable comments from seminar participants at the University of Pennsylvania. All remaining errors are our own. The views expressed in this paper are those of the authors and do not necessarily represent the views of the International Monetary Fund, its Management, or its Executive Directors. Sasaki gratefully acknowledges the research support provided by Brian and Charlotte Grove.}}

\author{Tobias Adrian\thanks{Tobias Adrian: Financial Counsellor and Director of the Monetary and Capital Markets Department, International Monetary Fund. Email: \texttt{TAdrian@imf.org}.}
\and
Yuya Sasaki\thanks{Yuya Sasaki: Brian and Charlotte Grove Chair and Professor of Economics, Vanderbilt University. Email: \texttt{yuya.sasaki@vanderbilt.edu}.}
\and 
Yulong Wang\thanks{Yulong Wang: Associate Professor of Economics, Lehigh University. Email: \texttt{yuw925@lehigh.edu}.}
}

\date{}
\maketitle
\begin{abstract}
\noindent The Growth-at-Risk (GaR) framework has garnered attention in recent econometric literature, yet current approaches implicitly assume a constant Pareto exponent. We introduce novel and robust econometrics to estimate the tails of GaR based on a rigorous theoretical framework and establish validity and effectiveness. Simulations demonstrate consistent outperformance relative to existing alternatives in terms of predictive accuracy. We perform a long-term GaR analysis that provides accurate and insightful predictions, effectively capturing financial anomalies better than current methods.  

\end{abstract}

\newpage
\section{Introduction}

Recent literature has seen a burgeoning interest in the study of extreme outcomes, exemplified by frameworks such as Financial Stability at Risk as measured by CoVaR \citep{adrian2016covar}, Employment at Risk \citep{ramey2018government}, Growth at Risk \citep{adrian2019vulnerable}, Inflation at Risk \citep{loria2022inflation}, 
and the Term Structure of Growth at Risk \citep{adrian2022term}, to name but a few.
Given the scarcity of data in the tails of outcome distributions, conventional econometric methods are often inadequate for these analyses. To address this challenge, researchers have developed novel econometric techniques tailored specifically to evaluate and quantify risks associated with such extreme outcomes.

To fix ideas, we focus on the Growth-at-Risk (GaR) framework of \citet{adrian2019vulnerable}.
Let $Y_{t+h}$ denote the average annual GDP growth rate between periods $t$ and $t+h$. 
Let $X_{t}$ denote a vector of observed controls. 
Researchers are interested in measuring the tail features of the conditional distribution of $Y_{t+h}$ given $X_t$.
For instance, a primary object of interest is the conditional quantile
\begin{align*}
Q_{Y_{t+h}|X_t}(\tau|x)
\end{align*}
of $Y_{t+h}$ given $X_t=x$ at extremely small $\tau$ or extremely large $\tau \in (0,1)$.
Furthermore, the expected shortfall and expected longrise are defined by 
\begin{align*}
SF_{Y_{t+h}|X_t}(\pi|x) =& \E[Y_{t+h}|Y_{t+h} \leq Q_{Y_{t+h}|X_t}(\pi|x), X_t=x]
\qquad\text{and}
\\
LR_{Y_{t+h}|X_t}(\pi|x) =& \E[Y_{t+h}|Y_{t+h} \geq Q_{Y_{t+h}|X_t}(1-\pi|x), X_t=x],
\end{align*}
respectively.

While the existing literature offers several valuable methods for analyzing these measures, it also has notable limitations. 
For example, if one begins with linear quantile regressions $Q_{Y_{t+h}|X_t}(\tau|x)=x'\beta(\tau)$ and extrapolates them to the tails using conditional skewed-t or Pareto-L\'evy distributions \citep{mandelbrot1960pareto} of $Y_{t+h}$ given $X_t=x$, then the resulting estimation strategy implicitly assumes a conditional Pareto exponent (equivalent to the degree of freedom of Student-t) that remains constant with respect to $x$. 
Consequently, the expected shortfall $SF_{Y_{t+h}|X_t}(\pi|x)$ and expected longrise $LR_{Y_{t+h}|X_t}(\pi|x)$ become proportional to the baseline linear quantile regression $Q_{Y_{t+h}|X_t}(\tau|x)=x'\beta(\tau)$.
In this paper, we will theoretically demonstrate these results.

Given the limitations of existing methods, we propose a more robust econometric approach for analyzing these parameters. 
Unlike existing methods, our approach does not imply a constant Pareto exponent with respect to $x$.
We provide a formal asymptotic theoretical foundation to validate our proposed method. 
Additionally, simulation studies based on data-generating processes that mimic the real-world data used in \citet{adrian2019vulnerable} demonstrate that our method consistently outperforms existing alternatives in terms of prediction accuracy.

We analyze GaR over the long history of the U.S. economy by combining the MacroHistory dataset with supplementary historical data.\footnote{\url{https://www.macrohistory.net/database/}} We demonstrate that our newly proposed method provides more informative GaR predictions, better capturing financial anomalies compared to existing methods. 
Specifically, our method's out-of-sample predicted quantiles contain the realized real GDP growth rates at frequencies close to the nominal coverage probabilities, outperforming the existing approach.

The remainder of this paper is structured as follows.
Section \ref{sec:existing} discusses the existing methods.
Section \ref{sec:novel} introduces our robust approach.
Section \ref{sec:asymptotic} develops the asymptotic theory.
Section \ref{sec:simulation} presents simulation evidence supporting our method. 
Section \ref{sec:history} applies our proposed method to study the GaR in the U.S. history.
Section \ref{sec:summary} concludes with discussions.
Detailed mathematical derivations and additional simulations are provided in the appendix.

\section{Existing Methods}\label{sec:existing}
\subsection{Review of the Existing Method}\label{sec:review_adrian}

We begin with reviewing a popular existing econometric methods. 
While we highlight the work of \citet{adrian2019vulnerable} due to its prominence, with other existing approaches exhibiting similar underlying characteristics. \citet{adrian2019vulnerable} propose the following estimation strategy.

\bigskip\noindent
{\bf Step 1:} 
First, estimate the linear quantile regression model $Q_{Y_{t+h}|X_t}(x) = x'\beta(\tau)$ by 
\begin{align*}
\hat\beta(\tau) = \arg\min_{\beta} \frac{1}{T}\sum_{t=1}^T \rho_\tau(Y_t - X_t'\beta)
\end{align*}
for each $\tau \in \mathcal{T} = \{0.05,0.25,0.75,0.95\}$,
where $\rho_\tau(u) = u \cdot (\tau - \mathbbm{1}\{u < 0\})$.
Write
$
\tilde Q_{Y_{t+h}|X_t}(\tau|x) = x'\hat\beta(\tau).
$

\bigskip\noindent
{\bf Step 2:}
Next, let $F_{Y_{t+h}|X_t}( \ \cdot \ ; \theta(x))$ denote the conditional CDF of the skewed-t distribution of $Y_{t+h}$ given $X_t=x$ with the parameters $\theta(x) = (\mu(x),\sigma(x),\alpha(x),v(x))'$ representing the location $\mu(x)$, scale $\sigma(x)$, skewness $\alpha(x)$, and degree of freedom $v(x)$, all of which are indexed by the conditioning variable $X_t=x$.
Estimate $\theta(x)$ by
\begin{align*}
\hat{\theta}(x) 
=
\arg \min_{\theta}
\sum_{\tau \in \mathcal{T}} \left( \tilde{Q}_{Y_{t+h}|X_t}(\tau|x) - F_{Y_{t+h}|X_t}^{-1}(\tau ;\theta) \right)^{2},
\end{align*}
i.e., fitting the skewed-t conditional distribution to the linear quantile regression estimates at $\tau \in \mathcal{T}$.

\bigskip\noindent
{\bf Step 3:}
If one wishes to proceed with estimating the expected shortfall $SF_{Y_{t+h}|X_t}(\pi|x)$ or the expected longrise $LR_{Y_{t+h}|X_t}(\pi|x)$ with tail probability $\pi$, then they can be estimated by extrapolating the estimated skewed-t distribution:
\begin{align*}
\widehat{SF}_{Y_{t+h}|X_t}(\pi|x) =& \frac{1}{\pi} \int_0^\pi F^{-1}_{Y_{t+h}|X_t}( \tau ; \hat\theta(x)) d\tau
\qquad\text{and}
\\
\widehat{LR}_{Y_{t+h}|X_t}(\pi|x) =& \frac{1}{\pi} \int_{1-\pi}^1 F^{-1}_{Y_{t+h}|X_t}( \tau ; \hat\theta(x)) d\tau.
\end{align*}

\subsection{Limitations of the Existing Methods}\label{sec:limitations}

\subsubsection{Constancy of $v(x)$ in $x$}

The first limitation of the existing method outlined in Section \ref{sec:review_adrian} is the implied constancy of degree of freedom $v(x)$ in $x$, as formally stated in the following proposition.

\begin{proposition}\label{prop:imposs}
Suppose that (i) $Q_{Y_{t+h}|X_t}\left( \tau |x\right) =x^{\prime }\beta\left( \tau \right)$ holds for $\tau \in [1-\varepsilon, 1)$ or $\tau \in (0, \varepsilon]$ for some $\varepsilon>0$ and (ii) $F_{Y_{t+h}|X_t}\left( \ \cdot \ |x\right)$ is a skewed-t distributed with degree of freedom $v\left( x\right)$.
Then $v(x)$ must be constant in $x$.
\end{proposition}

This result originates from \citet[][Proposition 2.1]{wang2013estimation}, who study the constancy of the tail index. 
We tailor their argument for the skewed-t distribution in our setup. 
A proof is found in Appendix \ref{sec:prop:imposs}.
This proposition establishes that the linear quantile regression model $Q_{Y_{t+h}|X_t}(x) = x'\beta(\tau)$ assumed in Step 1 of Section \ref{sec:review_adrian} and the skewed-t distribution of $F_{Y_{t+h}|X_t}\left( \ \cdot \ |x\right)$ assumed in Step 2 of Section \ref{sec:review_adrian} necessarily imply that degree of freedom $v(x) = \overline v$ is constant and does not depend on observed predictor $x$. 
In summary, while the procedure of \citet{adrian2019vulnerable} gives a tail index $v(x)$ that depends on $x$, their model implies that the tail in fact does not depend on $x$ but is a constant $v(x) = \overline v$.
Proposition \ref{prop:imposs} implies an internal inconsistency within the existing method: fitting the skewed-t distribution and using quantile regression are self-contradicting.  Specifically, quantile regression implies that the conditional quantile depends on $x$ through a linear location-and-scale transformation. Such location-and-scale model restricts that the tail index, which corresponds to the degree of freedom in the skewed-t distribution, to be constant. Therefore, fitting the skewed-t distribution automatically loses such dependence when using extreme quantiles. Since GaR precisely targets the extreme quantile and other tail features, losing the dependence of tail shape on $x$ could incur substantial misspecification bias.

\subsubsection{Constant Proportionality of $SF_{Y_{t+h}|X_t}(\cdot|x)$ and $LR_{Y_{t+h}|X_t}(\cdot|x)$ to $Q_{Y_{t+h}|X_t}(\cdot|x)$}

The second limitation of the existing method outlined in Section \ref{sec:review_adrian} is that it forces the expected shortfall $SF_{Y_{t+h}|X_t}(\pi|x)$  and the expected longrise $LR_{Y_{t+h}|X_t}(\pi|x)$ to be constant proportions of their threshold value $\pi$ regardless of the value of $x$.
Hence, there appears little point of making predictions based on $x$.

Since the mathematical analysis is symmetric, our presentation focuses on the expected longrise $LR_{Y_{t+h}|X_t}(\pi|x)$ in this section.

\begin{proposition}\label{prop:tce}
Suppose that the conditional distribution of $Y_{t+h}$ given $X_t=x$ is skewed-t with $v(x)$ degrees of freedom.
Then as $\pi \rightarrow 0$, 
\begin{equation*}
\frac{LR_{Y_{t+h}|X_t}\left( \pi|x\right) }{Q_{Y_{t+h}|X_t}(1-\pi|x)}\rightarrow
\left\{ 
\begin{array}{cl}
\frac{v (x)}{v (x)-1} & \text{if }v (x)>1 \\ 
\infty  & \text{otherwise.}
\end{array}
\right. 
\end{equation*}
\end{proposition}

A proof is found in Appendix \ref{sec:prop:tce}.
This proposition shows that the proportion of the expected longrise $LR_{Y_{t+h}|X_t}(\pi|x)$ to its threshold value $Q_{Y_{t+h}|X_t}(1-\pi|x)$ depends on only the degree of freedom $v(x)$, and does \textit{not} depend on the location $\mu(x)$, scale $\sigma(x)$, or skewness $\alpha(x)$.
In other words, the linear regression model $Q_{Y_{t+h}|X_t}(1-\pi|x) = x'\beta(1-\pi)$ must have already accounted for all the effects of the location $\mu(x)$, scale $\sigma(x)$, or skewness $\alpha(x)$ on the expected longrise.

When moment exists, it follows from combining Propositions \ref{prop:imposs}--\ref{prop:tce} that
\begin{align*}
LR_{Y_{t+h}|X_t}(\pi|x) \approx \frac{\overline v}{\overline v-1} \cdot x'\beta(1-\pi)
\end{align*}
holds for sufficiently small $\pi$, where the degrees of freedom $\overline v=v(x)$ is constant in $x$.
In other words, the expected longrise $LR_{Y_{t+h}|X_t}(\pi|x)$ is roughly a constant multiple of the linear quantile regression $Q_{Y_{t+h}|X_t}(1-\pi|x)=x'\beta(1-\pi)$ .

While the above discussion focuses on the expected longrise for brevity of exposition, an analogous conclusion holds also for the expected shortfall.
Namely, 
\begin{align*}
SF_{Y_{t+h}|X_t}(\pi|x) \approx \frac{\overline v}{\overline v-1} \cdot x'\beta(\pi)
\end{align*}
holds for sufficiently small $\pi$.

\section{Our New Econometric Method of Growth at Risk}\label{sec:novel}

To overcome the limitations of the existing method, we propose to consider a nonparametric family of conditional distributions $Y_{t+h}$ given $X_t=x$ that admit the Pareto tail approximation:
\begin{equation}\label{eq:pareto_approximation}
\mathbb{P}\left( Y_{t+h}>y|Y_{t+h}>y_{\min },X_{t}=x\right) \sim \left( \frac{y}{y_{\min }}\right) ^{-v\left( x\right) }
\end{equation}
for a sufficiently large $y_{\min }$,
where $v(x)$ denotes the tail exponent.
We emphasize that the skewed-t distribution employed by \citet{adrian2019vulnerable} satisfies this approximation condition \eqref{eq:pareto_approximation} with $v(x)$ corresponding to the degrees of freedom.
Besides the skewed-t distribution, it accommodates many parametric and nonparametric families \citep[e.g.,][Chapter 1]{dehaan2006extreme}.
Hence, this assumption \eqref{eq:pareto_approximation} is fairly weak and much weaker than the implicit assumptions made in the prior literature.

Furthermore, since the problem with the existing method was the constancy of $v(x)$ in $x$, we explicitly model its dependence on $x$ through
\begin{equation}\label{eq:v_exp}
v\left( x\right) =\exp \left( x^{\prime }\beta \right)
\end{equation}
for example.
The log-linear form has been explored by the statistics literature \citep[e.g.,][]{wang2009tail,wang2013estimation} with two major benefits: (i) it guarantees a positive tail index, and (ii) the tail index regression problem in Step 1 is strictly convex.  
We thus consider this functional form as a parsimonious and convenient specification.
Again, we want to emphasize that this model \eqref{eq:v_exp} is much more general than the implicit assumption imposed in the prior literature that $v(x)=\overline v$ is constant in $x$.

\subsection{The Estimation Procedure}\label{sec:new_method}

\subsubsection{ Algorithm for the Upper Tail}\label{sec:upper}

Given the asymmetry, we estimate the upper and lower tail features separately. 
Our proposed procedure for the upper tail is as follows. 
We denote the query point by $x_0$.

\bigskip\noindent
{\bf Step 1:} 
Run the tail index regression by solving
\begin{equation*}
\hat{\beta}=\arg \min_\beta \frac{1}{T}\sum_{t=1}^{T} \mathbbm{1}\{Y_{t+h} \geq y_{\min }\} \cdot \left\{ \exp (X_{t}^{\prime}\beta )\log \left( ( Y_{t+h} - \overline Y) / (y_{\min } - \overline Y) \right) -X_{t}^{\prime }\beta \right\} ,
\end{equation*}
where $\overline Y$ is the sample median of $\{Y_t\}_t$ and $y_{\min }$ is the 90\% empirical quantile of $\{Y_{t}\}_t$. 

\bigskip\noindent
{\bf Step 2:}
Use $\hat{\beta}$ to estimate
$
\hat v\left( x_0\right) =\exp ( x_0^{\prime }\hat \beta ).
$

\bigskip\noindent
{\bf Step 3:}
Estimate the $\tau$-th extreme quantile $Q_{Y_{t+h}|X_t}(\tau|x_0)$ in the upper tail quantiles by
\begin{equation*}
\hat{Q}^{\text{Upper}}_{Y_{t+h}|X_t}(\tau|x_0)= (y_{\min} - \overline Y) \cdot \left( \frac{1-\tau}{1-\hat F_{Y_{t+h}|X_t}(y_{\min}|x_0)}\right) ^{-1/\hat v(x_0) } + \overline Y,
\end{equation*}
where $\hat F_{Y_{t+h}|X_t}(y_{\min}|x_0)$ is a kernel estimator of $F_{Y_{t+h}|X_t}(y_{\min}|x_0)$ with bandwidth $b_T$. 

\bigskip\noindent
{\bf Step 4:}
Estimate the expected longrise $LR_{Y_{t+h}|X_t}(\tau|x_0)$ by
\begin{equation*}
\widehat{LR}_{Y_{t+h}|X}(\pi|x_0)= \left( \hat{Q}^{\text{Upper}}_{Y_{t+h}|X_t}(\tau|x_0) - \overline Y \right) \cdot \frac{\hat v\left( x_0\right) }{\hat v\left( x_0\right) -1} + \overline Y.
\end{equation*}

We provide some discussions. 
Step 1 is the tail index regression proposed by \citet{wang2009tail}. 
We modify their approach by subtracting the sample median $\overline{Y}$, which will not change the asymptotic distribution. 
The tail cutoff $y_{\min}$ serves as a tuning parameter, close in spirit to the bandwidth in kernel regression. 
We use 90\% quantile as a simple rule-of-thumb choice here and present a more sophisticated data-driven choice in Section \ref{sec:rule_of_thumb}. 
Once we obtain the tail index estimator $\hat{v}(x_0)$, we now take advantage of the Pareto tail approximation to estimate in the conditional extreme quantile as in Step 3. 
The expected longrise can be estimated analogously. 
\subsubsection{ Algorithm for the Lower Tail}

Our proposed procedure for the lower tail is same as before. 
We present it here for completeness and easy reference.

\bigskip\noindent
{\bf Step 1:} 
Run the tail index regression by solving
\begin{equation}\label{eq:tir}
\hat{\beta}=\arg \min_\beta \frac{1}{T}\sum_{t=1}^{T} \mathbbm{1}\{Y_{t+h} \leq y_{\max }\} \cdot \left\{ \exp (X_{t}^{\prime}\beta )\log \left( ( Y_{t+h} - \overline Y) / (y_{\max } - \overline Y) \right) -X_{t}^{\prime }\beta \right\} ,
\end{equation}
where $\overline Y$ is the sample median of $\{Y_t\}_t$ and $y_{\max }$ is the 10\% empirical quantile of $\{Y_{t}\}_t$. 

\bigskip\noindent
{\bf Step 2:}
Use $\hat{\beta}$ to estimate
$
\hat v\left( x_0\right) =\exp ( x_0^{\prime }\hat \beta ).
$

\bigskip\noindent
{\bf Step 3:}
Estimate the $\tau$-th extreme quantile $Q_{Y_{t+h}|X_t}(\tau|x_0)$ in the lower tail quantiles by
\begin{equation*}
\hat{Q}^{\text{Lower}}_{Y_{t+h}|X_t}(\tau|x_0) = (y_{\max} - \overline Y) \cdot \left( \frac{\tau}{\hat F_{Y_{t+h}|X_t}(y_{\max}|x_0)}\right) ^{-1/\hat v(x_0) } + \overline Y,
\end{equation*}
where $\hat F_{Y_{t+h}|X_t}(y_{\max}|x_0)$ is a kernel estimator of $F_{Y_{t+h}|X_t}(y_{\max}|x_0)$ with bandwidth $b_T$. 

\bigskip\noindent
{\bf Step 4:}
Estimate the expected shortfall $SF_{Y_{t+h}|X_t}(\tau|x_0)$ by
\begin{equation*}
\widehat{SF}_{Y_{t+h}|X_t}(\tau|x_0)= \left( \hat{Q}^{\text{Lower}}_{Y_{t+h}|X_t}(\tau|x_0) - \overline Y \right) \cdot \frac{\hat v\left( x_0\right) }{\hat v\left( x_0\right) -1} + \overline Y.
\end{equation*}

\subsection{Standard Errors}
\subsubsection{Standard Errors for the Upper Tail}\label{sec:se_upper}

We now provide estimators for the standard errors.
The standard error for $\hat{Q}^{\text{Upper}}_{Y_{t+h}|X_t}(\tau|x_0)$ is given by
$$
\frac{\sqrt{ \hat{\Sigma} _{F}\left(y_{\min},y_{\min}|x_{0}\right) } \cdot \hat{Q}^{\text{Upper}}_{Y_{t+h}|X_t}(\tau|x_0)}{\sqrt{Tb_T^{\text{dim}(X_t)}} \cdot \hat v \left( x_{0}\right)},
$$
where $\hat{\Sigma} _{F}\left(y_{\min},y_{\min}|x_{0}\right)$ denotes some consistent estimator for the asymptotic variance of the kernel estimator $\hat F_{Y_{t+h}|X_t}\left( y_{\min}|x_{0}\right)$, that is, 
$$
\frac{\int K(u)^2 du}{\hat{g}(x_0)} \hat{F}_{Y_{t+h}|X_t}\left( y_{\min}|x_{0}\right) \left( 1- \hat{F}_{Y_{t+h}|X_t}\left(y_{\min}|x_{0}\right) \right),
$$
where $K(\cdot)$ denotes the kernel function and $\hat{g}(x_0)$ denotes some estimator of the density of $X_t$ evaluated at $x_0$. 
Similarly, the standard error for $\widehat{LR}_{Y_{t+h}|X_t}(\tau|x)$ is given by
$$
\frac{\sqrt{ \hat{\Sigma} _{F}\left(y_{\min},y_{\min}|x_{0}\right) } \cdot \widehat{LR}_{Y_{t+h}|X_t}(\tau|x_0)}{\sqrt{Tb_T^{\text{dim}(X_t)}} \cdot (1 - \hat v \left( x_{0}\right))}.
$$

\subsubsection{Standard Errors for the Lower Tail}\label{sec:se_lower}

The standard error for $\hat{Q}^{\text{Lower}}_{Y_{t+h}|X_t}(\tau|x_0)$ is analogous to the previous case:
$$
\frac{\sqrt{ \hat{\Sigma} _{F}\left(y_{\max},y_{\max}|x_{0}\right) } \cdot \hat{Q}^{\text{Lower}}_{Y_{t+h}|X_t}(\tau|x_0)}{\sqrt{Tb_T^{\text{dim}(X_t)}} \cdot \hat v \left( x_{0}\right)}.
$$
Similarly, the standard error for $\widehat{SF}_{Y_{t+h}|X_t}(\tau|x)$ is given by
$$
\frac{\sqrt{ \hat{\Sigma} _{F}\left(y_{\max},y_{\max}|x_{0}\right) } \cdot \widehat{SF}_{Y_{t+h}|X_t}(\tau|x)}{\sqrt{Tb_T^{\text{dim}(X_t)}} \cdot (1 - \hat v \left( x_{0}\right))}.
$$

\subsection{Rule-of-Thumb Choice of the Threshold}\label{sec:rule_of_thumb}

The choice of $y_{\min}$ (and $y_{\max}$) is similar to that of the bandwidth in kernel regression. 
There is a large literature in studying the optimal choice of $y_{\min}$ in the case without the covariate.
See, \citet{danielsson2001using,danielsson2016tail}, among many other. 
In the case with covariate, we provide a data-driven method to choose the threshold, following \citet{wang2009tail}. 
For simple illustration, we focus on the right tail.

If $Y_{t+h}$ is Pareto distributed conditional on $X_t$ with the Pareto exponent being $\exp(X_t^{\prime }\beta)$, then $\exp(X_t^{\prime }\beta) \log(Y_{t+h})$ is standard exponential, and according $\exp(-\exp(X_t^{\prime }\beta) \log(Y_{t+h}))$ becomes uniformly distributed over $(0,1)$. 
Therefore, we aim to choose the threshold $y_{\min}$ such that the data above $y_{\min}$ satisfy the uniform distribution to the best extent. 
To this end, given any choice of $y_{\min}$, we first construct the estimator $\hat{\beta}$ and construct
\begin{equation*}
\hat{U}_t = \exp(-\exp(X_t^{\prime }\hat{\beta}) \log(Y_{t+h}/y_{\min})),
\end{equation*}
which should be approximately uniform. 
To select the optimal $y_{\min}$, we can compare the empirical distribution of $\hat{U}_t$ with the standard uniform by considering the discrepancy measure
\begin{equation*}
D_T (y_{\min}) = T_0^{-1}\sum_{t=1}^{T-h} \left( \hat{U}_t - \hat{F}_U(\hat{U}_t)   \right)^2 \times \mathbbm{1}[Y_{t+h}>y_{\min}],
\end{equation*}
where $T_0=\sum_{t=1}^{T-h} \mathbbm{1}[Y_{t+h}>y_{\min}]$ is the \textit{tail} sample size, and $ \hat{F}_U(\cdot)$ denotes the empirical distribution of $\hat{U}_t$. 
Accordingly, we can select the optimal choice of $y_{\min}$ to minimize $D_T(\cdot)$, i.e.,
\begin{equation}
\label{eq:optimal:ymin}
y^{*}_{\min} = \arg\min_{y_{\min}} D_T (y_{\min}). 
\end{equation}

\section{Asymptotic Theory}\label{sec:asymptotic}

We now derive the asymptotic distribution of our estimators to formally justify the procedure outlined in Section \ref{sec:novel}. 
For brevity, we focus on the case of upper tails presented in Sections \ref{sec:upper} and \ref{sec:se_upper}.
Symmetric arguments will provide counterpart results for the lower tails.

We make the following high-level assumptions.
\begin{description}
\item[Assumption P (Pareto-Type Tail)]
$\{Y_t,X_t\}$ is strictly stationary. The conditional CDF satisfies
\begin{equation*}
1-F_{Y_{t+h}|X_t}\left( y|x\right) =c_{1}(x)y^{- v(x)}\left( 1+c_{2}(x)y^{-\rho (x)}+r(y,x)\right),
\end{equation*}
where $v(x)=\exp (x^{\prime}\beta )$, $0<\inf_{x}c_{1}(x)\leq \sup_{x}c_{1}(x)<\infty $, $0<\sup_{x}\left\vert c_{2}(x)\right\vert <\infty $, $\inf_{x}\rho (x)>0$, and $\sup_{x}\left\vert r(y,x)\right\vert y^{-\rho(x)}\rightarrow 0$ as $y\rightarrow \infty $. The support of $X_{t}$ is compact.

\item[Assumption N (Nonparametric Estimation)]
There exists a sequence of constants $r_{T}\rightarrow 0$ such that
\begin{equation*}
T^{1/2}r_{T}\left( \hat{F}_{Y_{t+h}|X_t}\left( \ \cdot \ |x_{0}\right) -F_{Y_{t+h}|X_t}\left( \ \cdot \ |x_{0}\right) \right) \Rightarrow Z(\cdot),
\end{equation*}
where $Z$ is a Gaussian process with zero mean and covariance function $\Sigma _{F}$ satisfying $\Sigma _{F}\left(y,y|x_{0}\right) \sim c(x_0)F_{Y_{t+h}|X_t}\left( y|x_{0}\right) \left( 1-F_{Y_{t+h}|X_t}\left(y|x_{0}\right) \right) $ as $y\rightarrow \infty $ for some constant $c(x_0)>0$.

\item[Assumption T (Tail Index Regression)]
The tail index regression estimator \eqref{eq:tir} satisfies that
$\hat{\beta}-\beta=O_{p}( \mathbb{E} [ y_{\min }^{-v(X_{t})/2} ] ) $ as $y_{\min }\rightarrow \infty $ (and $T\rightarrow \infty $).

\item[Assumption R (Convergence Rates)] 
$y_{\min}$ and $r_T$ satisfy 
(i) $y_{\min }^{-v(x_{0})/2}\sqrt{T}r_{T}\rightarrow \infty $, 
(ii) $\left( \mathbb{E}\left[ y_{\min}^{-v(X_{t})}\right] \right) ^{-1/2}r_{T}\log \left( 1-\tau \right) \rightarrow 0$, and 
(iii) $y_{\min }^{-v(x_{0})\rho (x_{0})}\sqrt{T}r_{T}\rightarrow 0$ as $T\rightarrow \infty $.
\end{description}

Assumption P imposes a second-order conditional Pareto tail, with the term $y^{-\rho(x_0)}$ characterizing the deviation from the exact Pareto -- see also \citet{wang2009tail}. 
Assumption N requires that the nonparametric estimator $\hat{F}_{Y_{t+h}|X_t}\left( \cdot |x_{0}\right)$ uniformly converges to a Gaussian process with the variance-covariance proportional to $1-F_{Y_{t+h}|X_t}(\cdot|x_0)$ in the right tail. 
Specifically, this can be established for kernel estimation with $r_T = b^{\text{dim}(X)/2}_T$, where $b_T$ is the bandwidth. 
The constant $c(x_0)$ then becomes $\int K(u)^2 du /g(x_0)$ where $g(\cdot)$ denotes the density of $X$ and $K(\cdot)$ the kernel function \citep[e.g.,][]{cai2002regression,li2008nonparametric}.\footnote{These papers establish the pointwise convergence of the kernel conditional CDF estimator. Extension to the uniform convergence follows from the standard empirical process argument \citep[e.g.,][]{arcones1994central}. Such a result is available upon request. }
Assumption T describes the convergence of the tail index regression estimator.
This condition is established by \citet{nicolau2023tail} and \citet{wang2009tail} for the case without subtracting $\bar{Y}$. 
Suppose $\bar{Y}$ converges to the median of $Y$ at the rate root-$T$, subtracting $\bar{Y}$ from the extreme values of $Y$ makes no asymptotic effect.  

Assumption R imposes restrictions about the tuning parameters. 
Specifically, Assumptions R.(i) and R.(ii) hold if $r_T$ converges to zero neither too fast or too slow, reflecting the bias-variance trade-off in kernel estimation. 
Assumption R.(iii) imposes some restriction about the Pareto tail approximation. 
Recall that a larger $\rho(x_0)$ implies a better Pareto tail approximation. 
Then, this condition essentially requires a sufficiently well approximation by imposing $\rho(x_0)>1/2$. 
For instance, the Student-t distribution satisfies this condition since $\rho(x_0)=2$. 
As a consequence, the convergence rate of the Pareto-tail estimation is faster than that of the kernel estimation. 
It follows that the asymptotic variance of our proposed estimator becomes dominated by that of the nonparametric estimation.

The following theorem establishes the limiting distributions of our proposed estimators.

\begin{theorem}\label{thm:asym}
Suppose Assumptions P, N, T, and R hold. Then,
\begin{equation*}
\left( T^{1/2}r_{T}\right) \Sigma _{Q}^{-1/2}\left( \frac{\hat{Q}^{\text{Upper}}_{Y_{t+h}|X_t}(\tau|x_0)}{Q_{Y_{t+h}|X_t}(\tau |x_0)}-1\right) \overset{d}{\rightarrow }\mathcal{N}\left(0,1\right)
\end{equation*}
holds as $T\rightarrow \infty $, where
\begin{equation*}
\Sigma _{Q}=\frac{\Sigma _{F}\left( y_{\min },y_{\min }|x_{0}\right) }{v \left( x_{0}\right)^{2}}.
\end{equation*}
In addition, 
\begin{equation*}
\left( T^{1/2}r_{T}\right) \Sigma _{T}^{-1/2}\left( \frac{\widehat{LR}_{Y_{t+h}|X_t}(\tau|x_0)}{LR_{Y_{t+h}|X_t}(\tau |x_0)}-1\right) \overset{d}{\rightarrow }\mathcal{N}\left(0,1\right) ,
\end{equation*}
holds as $T\rightarrow \infty $, where
\begin{equation*}
\Sigma _{T}=\frac{\Sigma _{F}\left( y_{\min },y_{\min }|x_{0}\right) }{\left( v \left( x_{0}\right) -1\right) ^{2}}.
\end{equation*}
\end{theorem}

\section{Simulation Comparisons in Prediction Accuracy}\label{sec:simulation}

In this section, we use simulation studies to contrast the performance of the existing method (Section \ref{sec:review_adrian}) and our proposed novel method (Section \ref{sec:new_method}) in the context of designs emulating the real data.

\subsection{Data Generating Designs}\label{sec:designs}

We develop two data generating designs based on the real data used by \citet{adrian2019vulnerable}.
These two designs are tailored to quarter-ahead predictions (Section \ref{sec:simu:quater}) and year-ahead predictions (Section \ref{sec:simu:year}) of GDP growth rates, based on the current GDP growth rate and a financial indicator.

\subsubsection{Design Based on Quarter-Ahead Predictions}\label{sec:simu:quater}

Our first data-generating process is as follows.
We generate $\{X_t\}_{t=1}^T$ independently by
\begin{align*}
X_t
=
\left(\begin{array}{c}
1 \\ X_{t1} \\ X_{t2}
\end{array}\right)
\text{ with }
\left(\begin{array}{c}
X_{t1} \\ X_{t2}
\end{array}\right)
\sim
\mathcal{T}\left(
\left(\begin{array}{c}
2.732 \\ 0.007
\end{array}\right),
\left(\begin{array}{cc}
10.671 & -1.152 \\ -1.152 & 0.972
\end{array}\right),
\left(\begin{array}{c}
6.360 \\ 7.064
\end{array}\right)\right),
\end{align*}
where $\mathcal{T}(m_X, S_X, d_X)$ denotes the joint distribution with marginal Student-t distributions having $d_X$ degrees of freedom, mean vector $m_X$, and covariance matrix $S_X$, with dependence structure governed by a Gaussian copula.
Note that this specification requires the degrees of freedom to be greater than 2, which is satisfied with the provided parameter values, 6.360 and 7.064.

Given $X_t$, we in turn generate $Y_t$ from the skewed-t distribution with parameters parameters $\theta(X_t) = (\mu(X_t),\sigma(X_t),\alpha(X_t),v(X_t))'$ given by
\begin{align*}
&
\mu(X_t) = 2.053 - 0.341X_{t1} - 1.678X_{t2},
&&
\sigma(X_t) = \exp(0.925 + 0.085X_{t1} + 0.437X_{t2}),
\\
&
\alpha(X_t) = -0.710 + 0.763X_{t1} - 1.218X_{t2},
&&
v(X_t) = \exp(2.848 -0.162X_{t1} + 0.303X_{t2}).
\end{align*}
These parameter values are obtained by fitting the known model of the conditional skewed-t distribution with the real data used by \citet{adrian2019vulnerable} for quarter-ahead predictions.

\subsubsection{Design Based on Year-Ahead Predictions}\label{sec:simu:year}

Our second data-generating process is as follows.
We generate $\{X_t\}_{t=1}^T$ independently by
\begin{align*}
X_t
=
\left(\begin{array}{c}
1 \\ X_{t1} \\ X_{t2}
\end{array}\right)
\text{ with }
\left(\begin{array}{c}
X_{t1} \\ X_{t2}
\end{array}\right)
\sim
\mathcal{T}\left(
\left(\begin{array}{c}
2.761 \\ 0.018
\end{array}\right),
\left(\begin{array}{cc}
10.806 & -1.193 \\ -1.193 & 0.981
\end{array}\right),
\left(\begin{array}{c}
14.216 \\ 7.685
\end{array}\right)\right),
\end{align*}
where $\mathcal{T}(m_X, S_X, d_X)$ was defined in Section \ref{sec:simu:quater}.
Given $X_t$, we in turn generate $Y_t$ from the skewed-t distribution with parameters parameters $\theta(X_t) = (\mu(X_t),\sigma(X_t),\alpha(X_t),v(X_t))'$ given by
\begin{align*}
&
\mu(X_t) = 2.301 - 0.107X_{t1} -0.289X_{t2},
&&
\sigma(X_t) = \exp(0.642 + 0.0589X_{t1} + 0.224X_{t2}),
\\
&
\alpha(X_t) = 1.019 + 0.087X_{t1} -0.668X_{t2},
&&
v(X_t) = \exp(1.214 + 0.115X_{t1} + 0.340X_{t2}).
\end{align*}
These parameter values are obtained by fitting the known model of the conditional skewed-t distribution with the real data used by \citet{adrian2019vulnerable} for year-ahead predictions.

\subsubsection{Remarks about the Data-Generating Designs}

Both of the above two data generating designs are developed in favor of the existing method reviewed in Section \ref{sec:review_adrian} in the sense that the conditional skewed-t distribution conforms with their parametric assumption.
On the other hand, the fact that degrees of freedom $v(X_t)$ is a non-constant function of $X_t$ induces a self-contradiction for this existing method, in light of the conclusion from our Proposition \ref{prop:imposs}.

\subsection{Simulation Results}

\begin{figure}[tbp]
    \centering
    \includegraphics[width=0.49\textwidth]{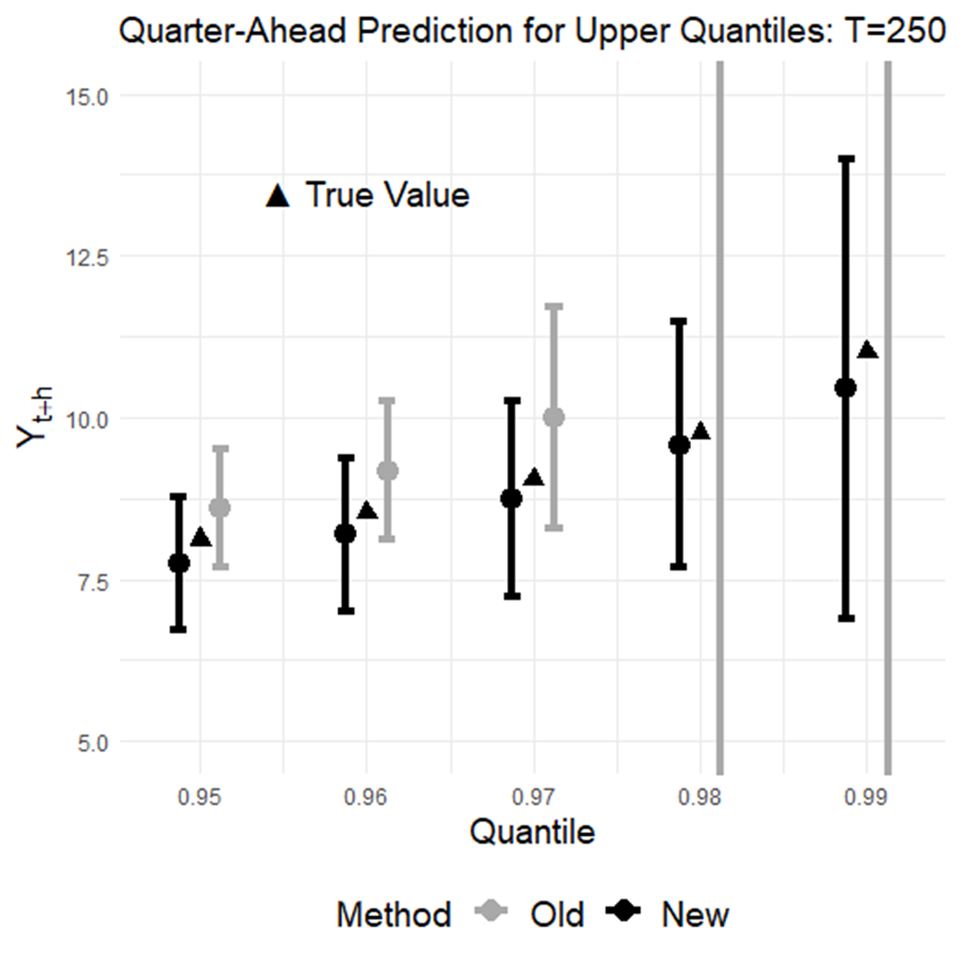}
    \includegraphics[width=0.49\textwidth]{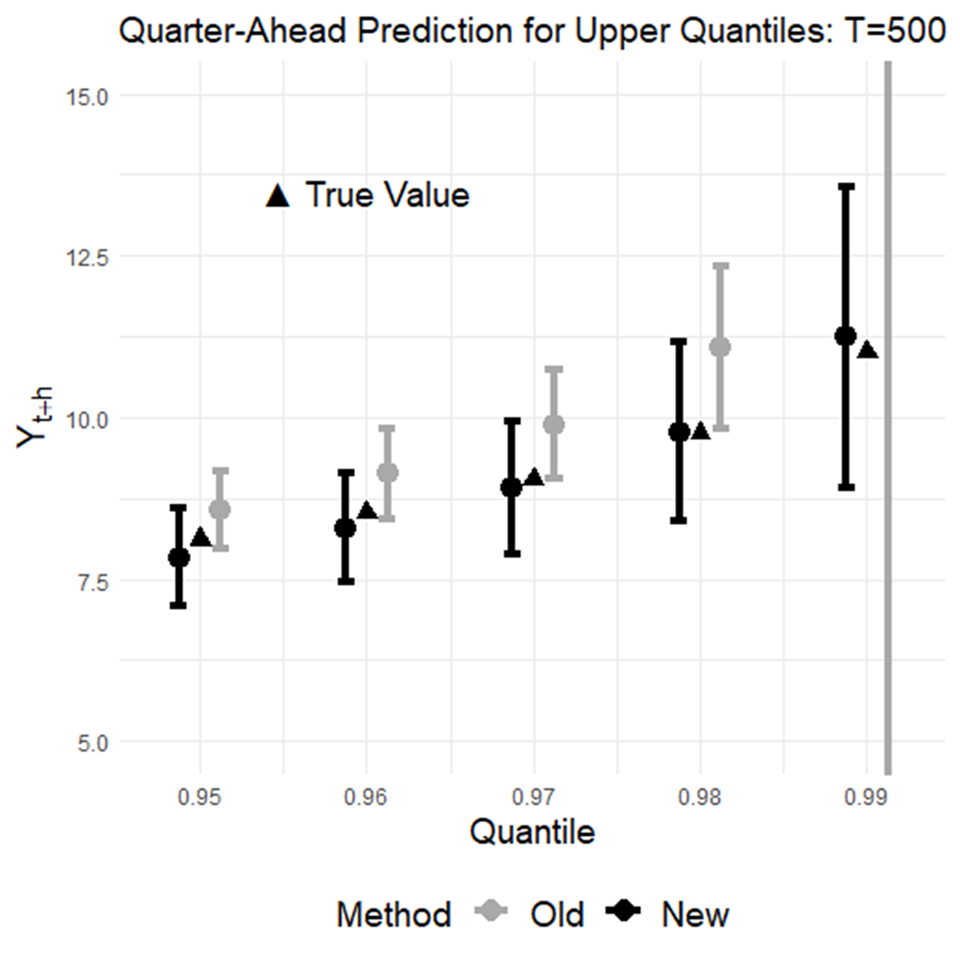}
    \\${}$\\
    \includegraphics[width=0.49\textwidth]{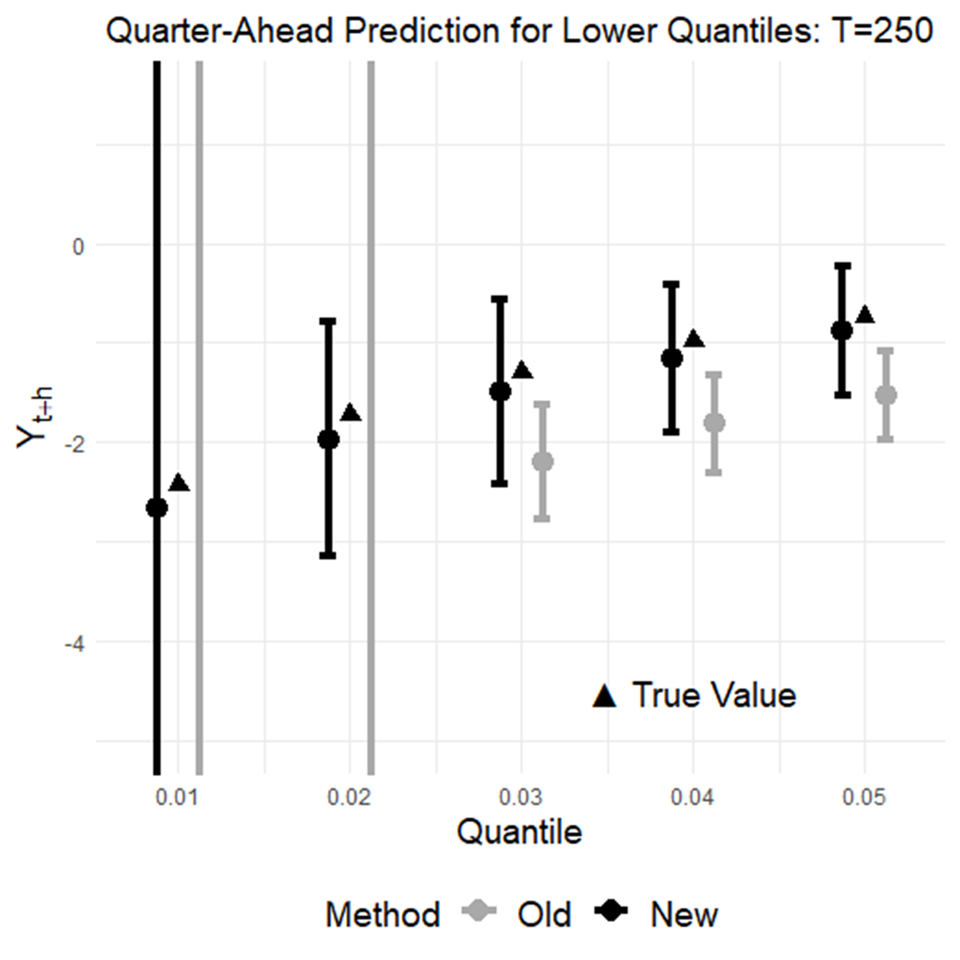}
    \includegraphics[width=0.49\textwidth]{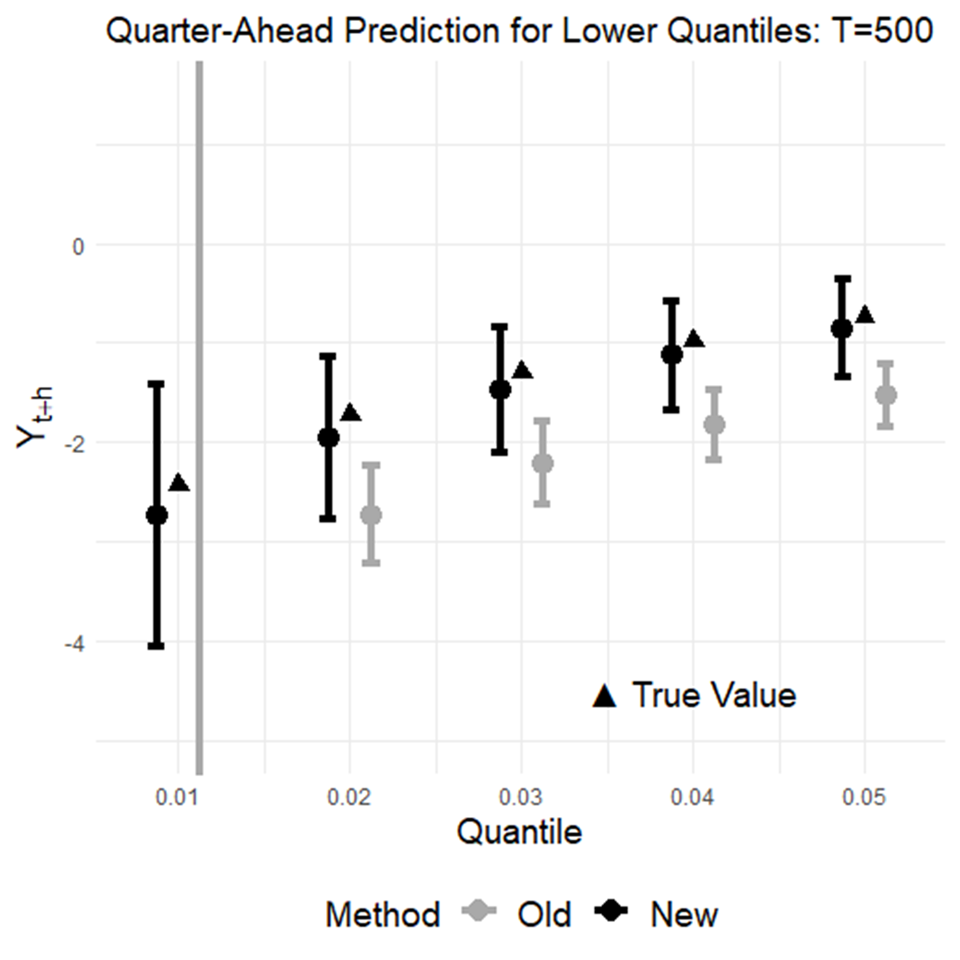}
    \caption{Simulation results comparing the performance of the proposed method (in black) and the existing method (in gray) for quarter-ahead predictions of growths. The predictions are conditional on the average values of $(X_{t1},X_{t2})= x_0 :=(2.732,0.007)$. The upper panels present results for the upper tail ($\tau \in \{0.95,0.96,0.97,098,0.99\}$), while the lower panels show results for the lower tail ($\tau \in \{0.01,0.02,0.03,0.04,0.05\}$). Dots represent simulation averages, bars represent the Gaussian interquartile ranges, and triangles denote the true values. The left column illustrates results for $T=250$, and the right column illustrates results for $T=500$. The results are based on 2,500 Monte Carlo iterations.}
    \label{fig:quarter_ahead}
\end{figure}

\begin{figure}[tbp]
    \centering
    \includegraphics[width=0.49\textwidth]{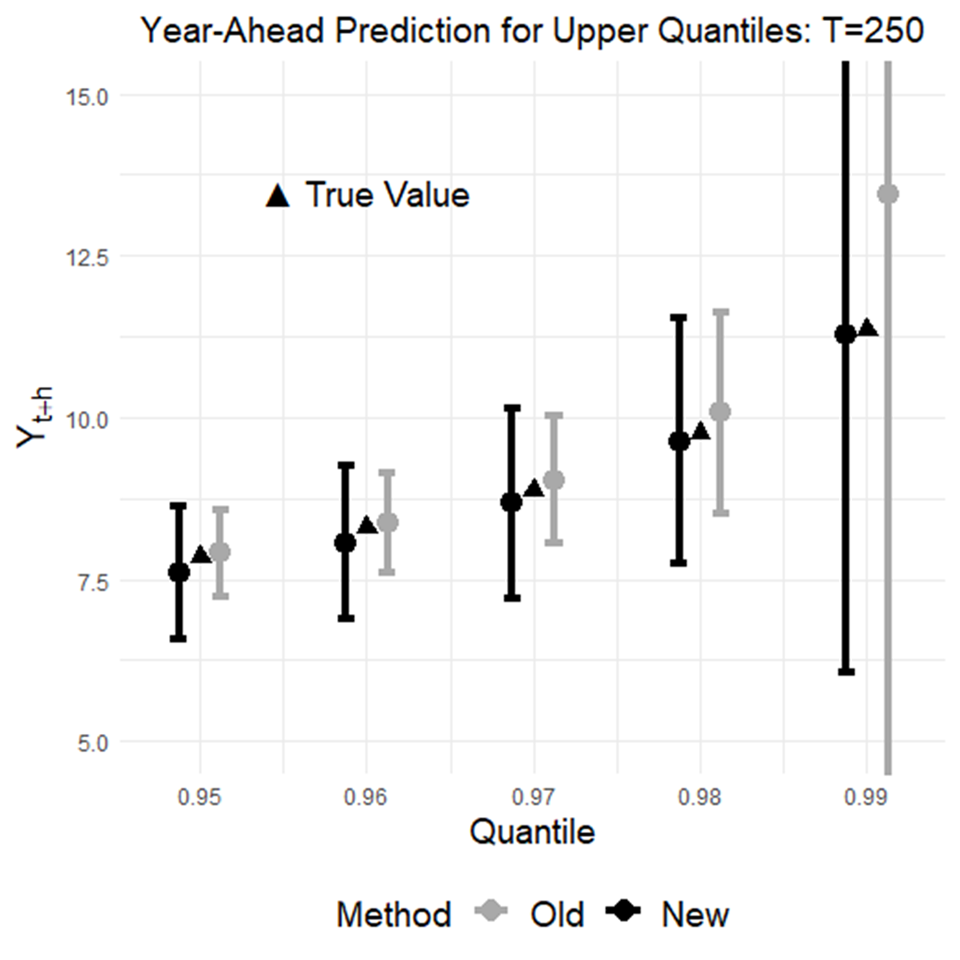}
    \includegraphics[width=0.49\textwidth]{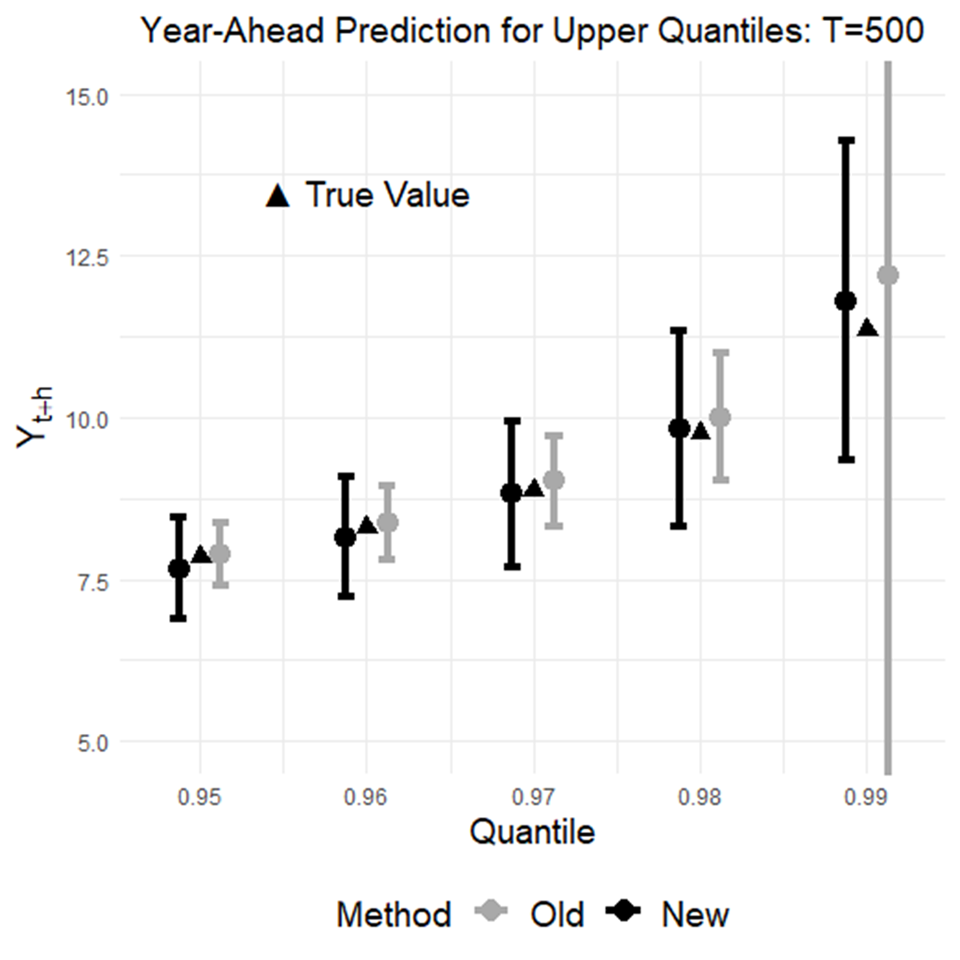}
    \\${}$\\
    \includegraphics[width=0.49\textwidth]{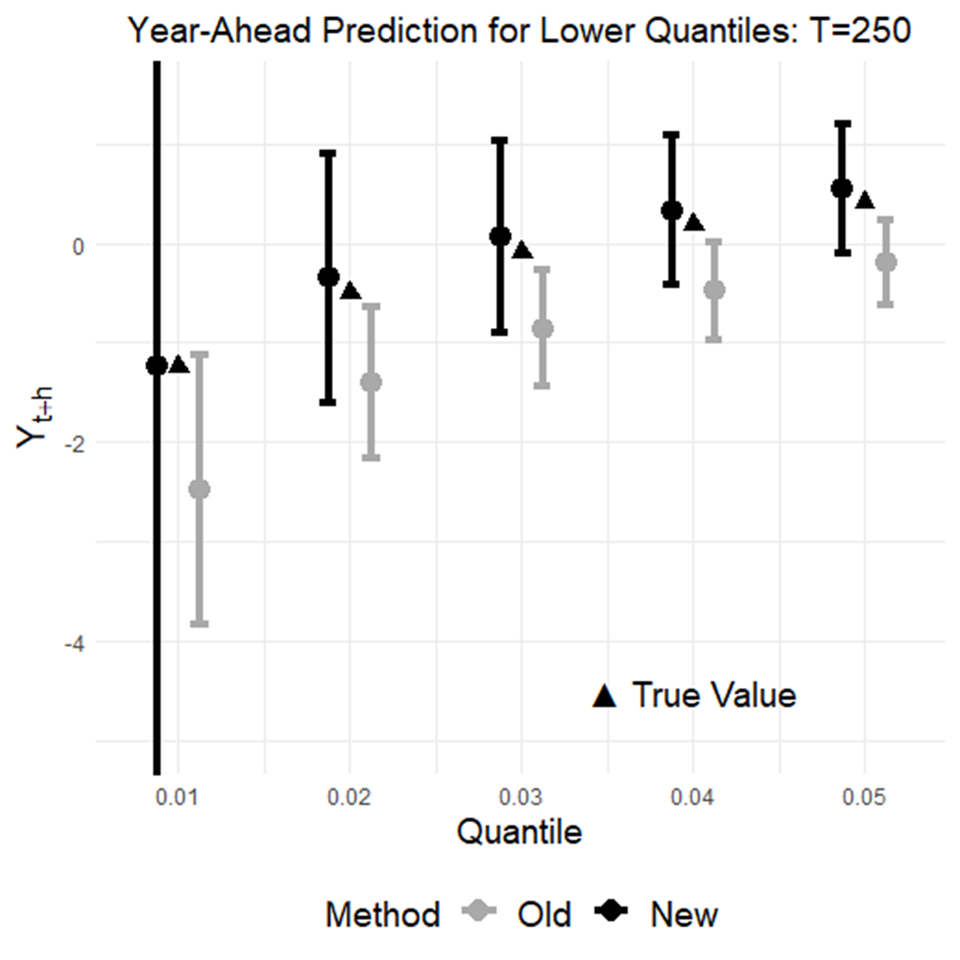}
    \includegraphics[width=0.49\textwidth]{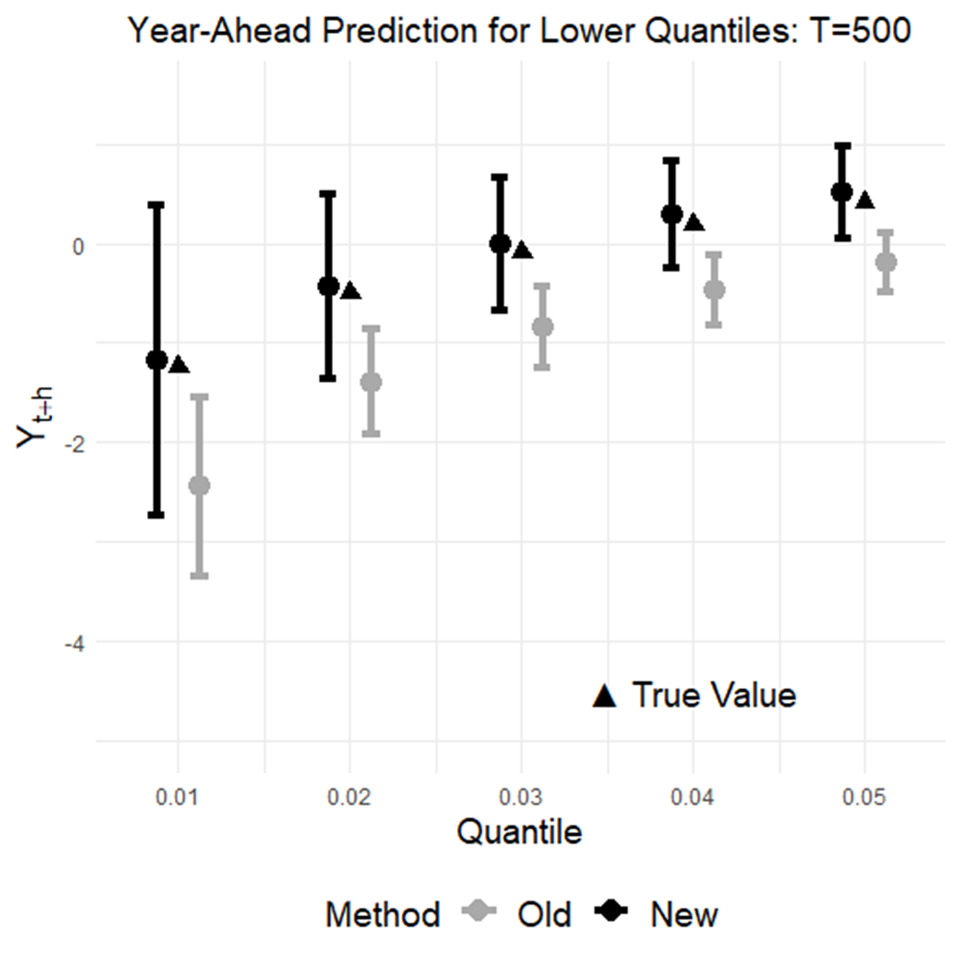}
    \caption{Simulation results comparing the performance of the proposed method (in black) and the existing method (in gray) for year-ahead predictions of growths. The predictions are conditional on the average values $(X_{t1},X_{t2})=x_0 :=(2.761,0.018)$. The upper panels present results for the upper tail ($\tau \in \{0.95,0.96,0.97,098,0.99\}$), while the lower panels show results for the lower tail ($\tau \in \{0.01,0.02,0.03,0.04,0.05\}$). Dots represent simulation averages, bars represent the Gaussian interquartile ranges, and triangles denote the true values. The left column illustrates results for $T=250$, and the right column illustrates results for $T=500$. The results are based on 2,500 Monte Carlo iterations.}
    \label{fig:year_ahead}
\end{figure}

Figures \ref{fig:quarter_ahead} and \ref{fig:year_ahead} illustrate the simulation results for quarter-ahead predictions ($h=1$) and year-ahead predictions ($h=4$), respectively.
The predictions $Q_{Y_{t+h}|X_t}(\tau|x_0)$ of $Y_{t+h}$ are conditional on the average values of $X_t$, i.e., $(X_{t1},X_{t2})= x_0 :=(2.732,0.007)$ in Figure \ref{fig:quarter_ahead} and $(X_{t1},X_{t2})= x_0 :=(2.761,0.018)$ in Figure \ref{fig:year_ahead}.
In each figure, the upper panels display results for the upper tail ($\tau \in {0.95, 0.96, 0.97, 0.98, 0.99}$), while the lower panels show results for the lower tail ($\tau \in {0.01, 0.02, 0.03, 0.04, 0.05}$).
Each panel includes simulation averages (represented by dots) and Gaussian interquartile ranges (shown as bars) of the estimates for both our proposed method (`New' in black), described in Section \ref{sec:new_method} and the existing method (`Old' in gray), reviewed in Section \ref{sec:review_adrian}.
Triangles mark the true values. 
Within each figure, the left column illustrates results for a sample size of $T=150$, and the right column illustrates results for a sample size of $T=300$.

First, we examine the quarter-ahead prediction results for the upper tail, displayed in the upper panels of Figure~\ref{fig:quarter_ahead}. In these panels, the average predicted values from the `New' method (represented by dots) closely align with the true values (represented by triangles) across all quantiles $\tau \in \{0.95, \ldots, 0.99\}$. In contrast, the predictions from the `Old' method deviate increasingly from the true values as the quantile level becomes more extreme. Moreover, at these extreme quantiles ($\tau \in \{0.98,0.99\}$ when $T=250$ and $\tau = 0.99$ when $T=500$), the `Old' method yields loose and uninformative predictions, as evidenced by the exploding interquartile ranges, whereas the `New' method produces stable and informative predictions.
Next, we consider the results for the lower tail, presented in the lower panels of Figure~\ref{fig:quarter_ahead}. Once again, the average predicted values from the `New' method closely align with the true values across all quantiles $\tau \in \{0.01, \ldots, 0.05\}$. The deviation of the predictions produced by the `Old' method is even more pronounced, and the interquartile ranges of its predicted values fail to contain the true value for quantiles $\tau \in \{0.03, 0.04, 0.05\}$ when $T = 250$, and for $\tau \in \{0.02, 0.03, 0.04, 0.05\}$ when $T = 500$.

Now, let us examine the year-ahead prediction results for the upper tail, shown in the upper panels of Figure~\ref{fig:year_ahead}. In these panels, the average predictions produced by both the `New' and `Old' methods are reasonably close to the true values. Nonetheless, the `Old' method tends to yield more biases and looser predictions, as indicated by the longer interquartile ranges, for the most extreme quantile $\tau = 0.99$. 
Turning to the lower tail results, presented in the lower panels of Figure~\ref{fig:year_ahead}, we observe a clear and consistent downward bias in the estimates produced by the `Old' method, whereas the `New' method remains closely aligned with the true values. The interquartile range of the predicted values produced by the `Old' method consistently fails to contain the true value for each of the quantiles $\tau \in \{0.01, \ldots, 0.05\}$ when $T = 500$. These results underscore the superior performance of the `New' method relative to the `Old' one.

\begin{figure}[tbp]
    \centering
    \includegraphics[width=0.49\textwidth]{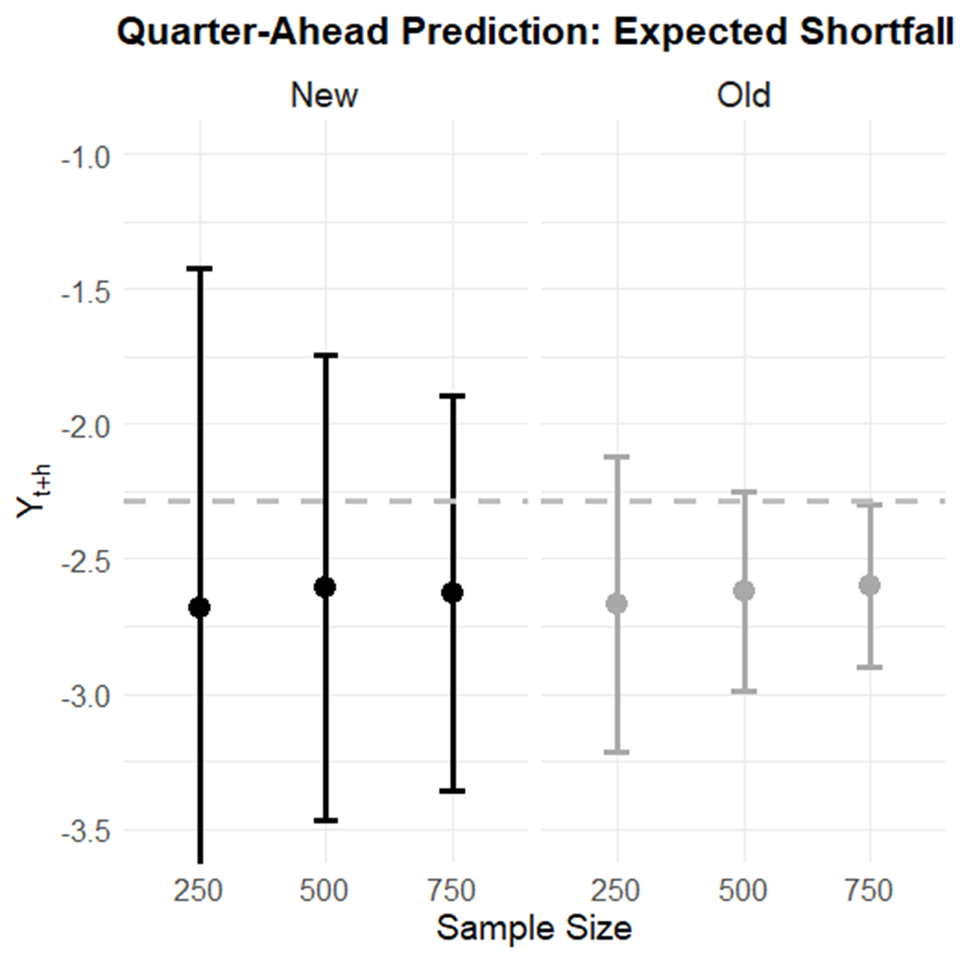}
    \includegraphics[width=0.49\textwidth]{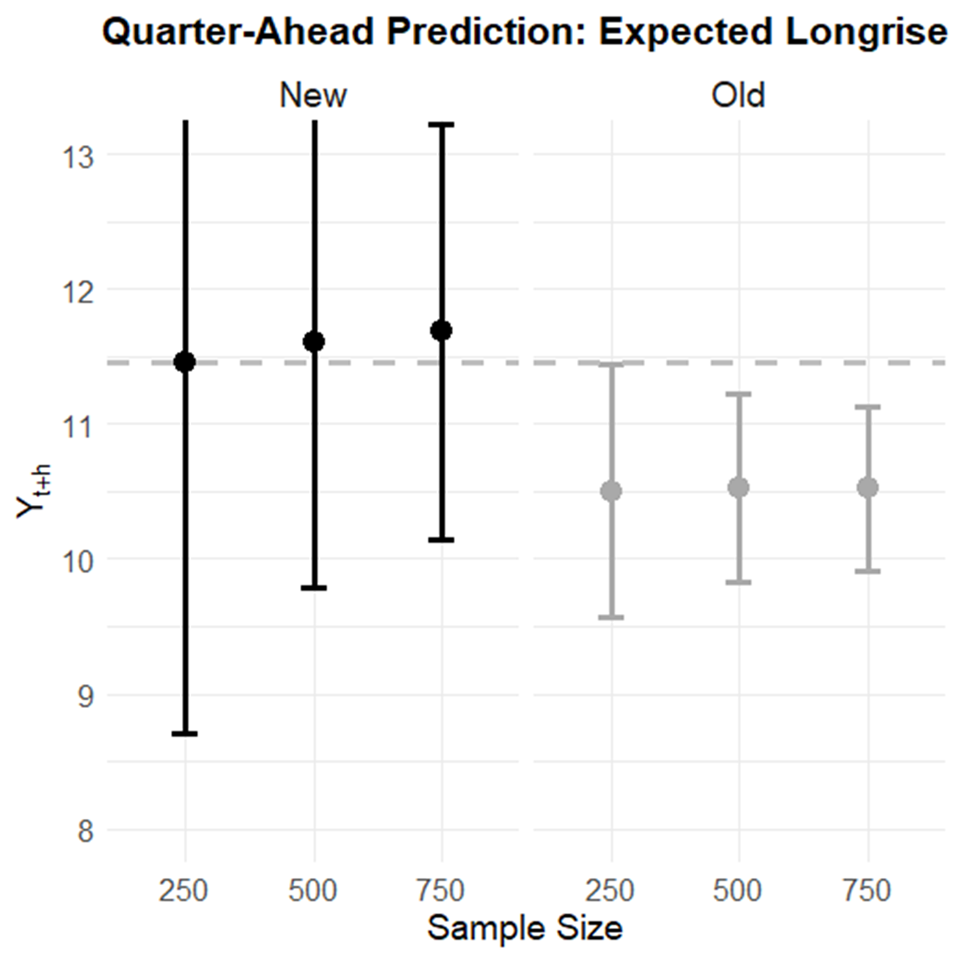}
    \\${}$\\
    \includegraphics[width=0.49\textwidth]{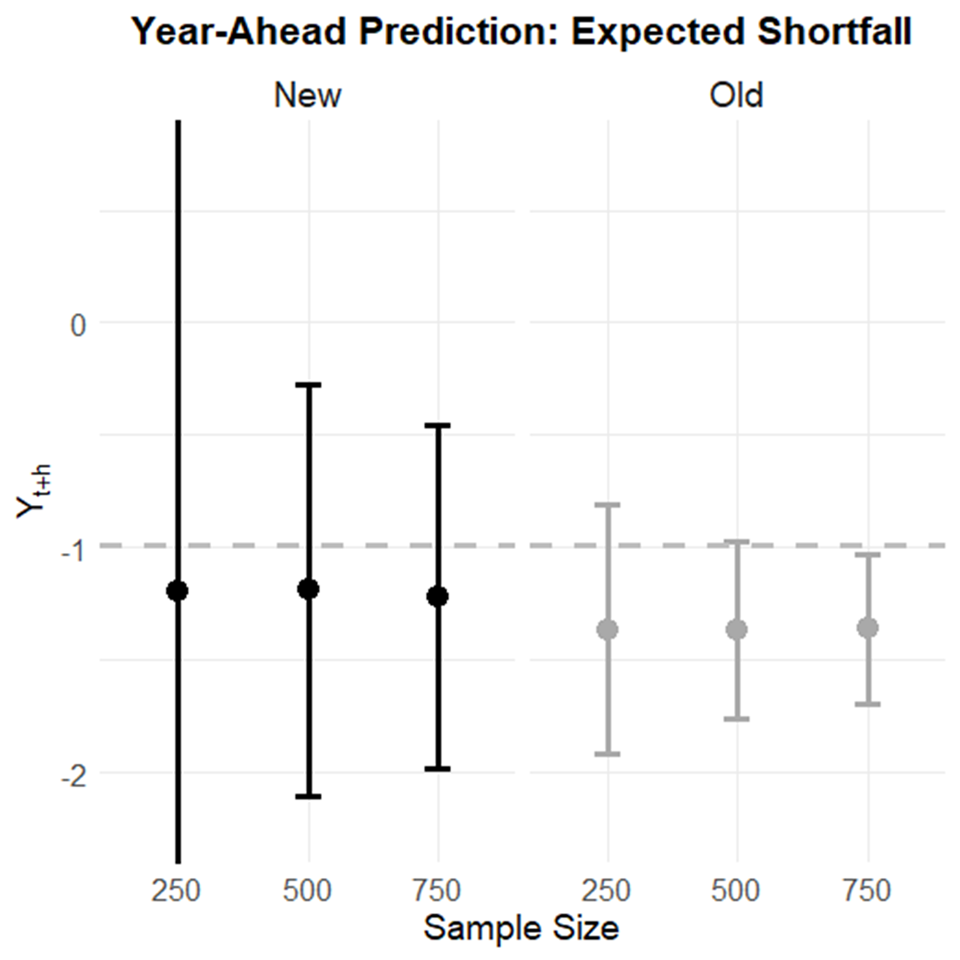}
    \includegraphics[width=0.49\textwidth]{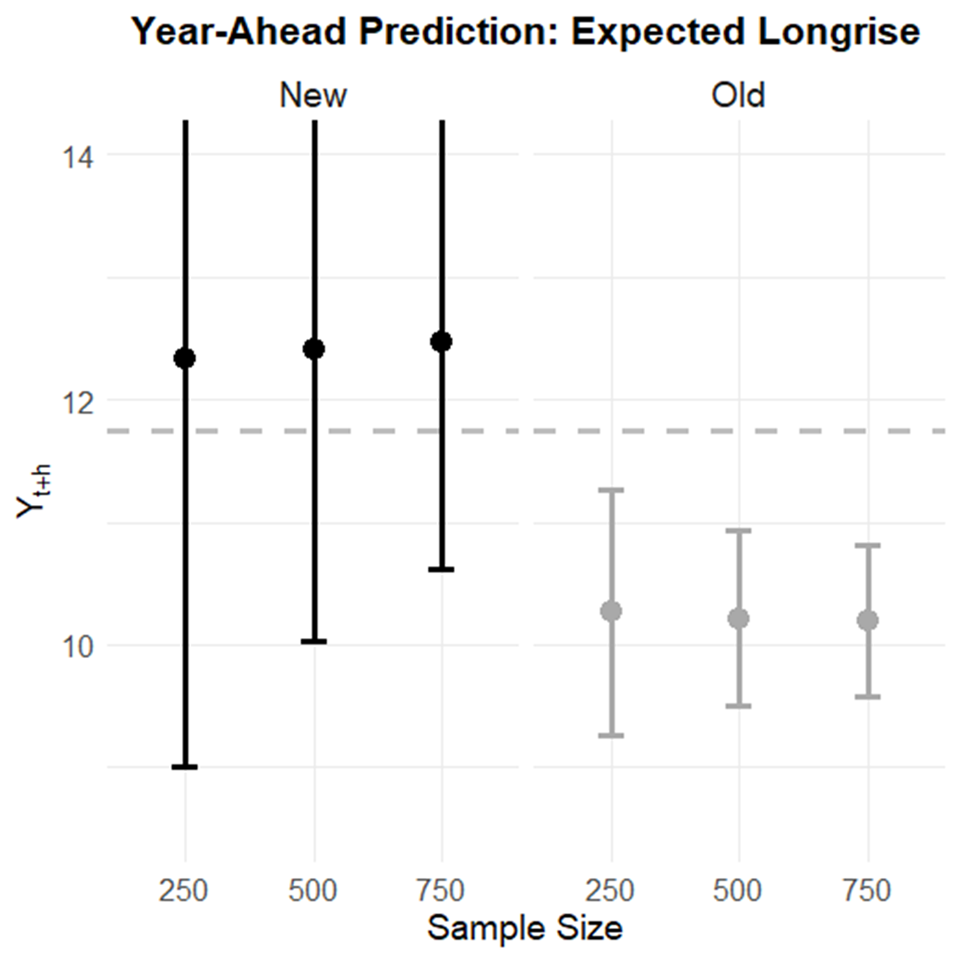}
    \caption{Simulation results comparing the performance of the proposed method (in black) and the existing method (in gray) for the expected shortfall $SF_{Y_{t+h}|X_t}(\pi|x_0)$ (left) and the expected longrise $LR_{Y_{t+h}|X_t}(\pi|x_0)$ (right) for $\pi=0.05$.  The upper panels present results for quarter-ahead predictions, while the lower panels show results for year-ahead predictions. The predictions are conditional on the average values of $(X_{t1},X_{t2})=x_0 :=(2.732,0.007)$ for the upper panels and $(X_{t1},X_{t2})=x_0 :=(2.761,0.018)$ for the lower panels. Dots represent simulation averages, bars represent the Gaussian interquartile ranges, and the dashed horizontal lines denote the true values. The results are displayed for each of the sample sizes $T=\{250,500,750\}$. The results are based on 2,500 Monte Carlo iterations.}
    \label{fig:SF_LR}
\end{figure}

Figure \ref{fig:SF_LR} illustrates the simulations results for the expected shortfall $SF_{Y_{t+h}|X_t}(\pi|x_0)$ (left) and the expected longrise $LR_{Y_{t+h}|X_t}(\pi|x_0)$ (right) for $\pi=0.05$.
The upper panels display results for quarter-ahead predictions, while the lower panels show results for year-ahead predictions. Each panel includes the simulation averages (represented by dots) and Gaussian interquartile ranges (shown as bars) of the estimates for both our proposed method (`New' in black) detailed in Section \ref{sec:new_method} and the existing method (`Old' in gray) discussed in Section \ref{sec:review_adrian}. 
The predictions are conditional on the average values of $(X_{t1},X_{t2})=x_0 :=(2.732,0.007)$ for the upper panels and $(X_{t1},X_{t2})=x_0 :=(2.761,0.018)$ for the lower panels.
The dashed horizontal lines represent the true values. 
Results are presented for each sample size $T \in \{250, 500, 750\}$.

We first examine the expected shortfall $SF_{Y_{t+h}\mid X_t}(\pi|x_0)$ displayed in the left column. For the quarter-ahead prediction (top panel), the magnitude of the bias is similar between the `New' and `Old' methods, but the interquartile range of the `Old' method fails to contain the true value when $T = 750$ due to its tightness. For the year-ahead prediction (bottom panel), the `Old' method exhibits roughly twice the downward bias of the `New' method, and again its interquartile range fails to contain the true value when $T = 750$.

Now let us consider the expected longrise $LR_{Y_{t+h}\mid X_t}(\pi|x_0)$ shown in the right column of the figure. For the quarter-ahead predictions (top panel), the `New' method is nearly unbiased, whereas the `Old' method exhibits a substantial downward bias. For the year-ahead predictions (bottom panel), similar patterns emerge, but with larger magnitudes. In both prediction horizons, the interquartile range of the predictions produced by the `Old' method fails to contain the true value for every sample size $T \in \{250,500,750\}$ considered in these studies.

\subsection{Summary of Simulation Comparisons}

The simulation results confirm that our proposed `New' method (Section \ref{sec:new_method}) outperforms the existing `Old' method (Section \ref{sec:review_adrian}) in predicting extreme growth. 
We developed data-generating designs based on real-world data to evaluate quarter-ahead and year-ahead predictions -- see Section \ref{sec:designs}. 
Notably, our designs incorporate a conditional skewed-t distribution, aligning with the parametric distributional assumptions implicitly used by the `Old' method, while our `New' method makes no such assumptions. 
It is therefore noteworthy that, even with its flexible distributional assumptions, the `New' method achieves superior predictive performance compared to the `Old' method.

\subsection{Additional Simulations}\label{sec:simulations_comment_additional}

We conduct an extensive array of additional simulation exercises beyond those presented above. The details are provided in Appendix~\ref{sec:additional_simulation}, which contains three sets of simulations.

First, while the simulation results presented in the main text rely on the data-driven choice of $y_{\min}$ and $y_{\max}$ proposed in Section~\ref{sec:rule_of_thumb}, Appendix~\ref{sec:additional_simulation:fixed} examines the performance of the `New' method under fixed choices of $y_{\min}$ and $y_{\max}$. Specifically, in that appendix, $y_{\min}$ and $y_{\max}$ are set simply to the 90\% and 10\% empirical quantiles of $\{Y_t\}_t$. The resulting performance is broadly similar to the baseline results obtained under the data-driven choices presented in the main text, with the data-driven choice performing slightly better in the more extreme quantiles.
We recommend using the data-driven choice for greater robustness.

Second, the current data-generating design entails a tail exponent $v(X_t)$ that is not constant in $X_t$, and thus does not accommodate the `Old' method due to the limitations discussed in Section~\ref{sec:limitations}. A natural question, then, is what happens to the simulation results when $v(X_t)$ is instead designed to be constant in $X_t$. Appendix~\ref{sec:additional_simulation:constant} presents simulations under this constant tail exponent model.
As expected, the performance of the `Old' method improves relative to the baseline design in which the tail exponent $v(X_t)$ varies with $X_t$. However, our proposed `New' method continues to substantially outperform the `Old' method even under this more favorable setting. Hence, we recommend using the more robust `New' method even if the researcher knows that the tail exponent is constant.

Third, the current data-generating design is based on a tail exponent model $v(X_t)$ that is correctly specified for the `New' method. A natural question, then, is how robust the `New' method remains under misspecification of this tail exponent model. Appendix~\ref{sec:additional_simulation:misspecification} investigates this issue by experimenting with misspecified tail exponent models. The results show that the `New' method remains quite robust against a moderate degree of misspecification and, in particular, continues to outperform the `Old' method.
With this said, it would be desirable to develop a more flexible semiparametric model for the tail exponent, and we leave this extension for future research.

\section{Predicting Growth at Risk in the U.S. History}\label{sec:history}

Having demonstrated the superior performance of our proposed method in predicting extreme growth events, we now apply this novel approach to a long-term U.S. macro-financial dataset spanning 1893--2016. The dataset is obtained from \citet{Gächter_Hasler_Huber_2025} stretching from 1893Q1 to 2016Q4 for the US and including annualized real GDP growth, a financial stress indicator, the 3-year
average growth rate of the credit-to-GDP ratio, and the 3-year average growth rate of real house
prices. The financial stress indicators is a historical newspaper-based financial stress indicator.

We perform out-of-sample predictions for the year-ahead growth rate $Y_{t+4}$, using the current growth rate $X_{t,1} (= Y_{t})$ and the current financial condition index (FCI) $X_{t,2}$ as predictors. 
Using the annual cumulative growth ($h = 4$) instead of the quarter-to-quarter growth serves primarily a communicative purpose, as central banks and statistical agencies typically report growth in annual terms. Second, a quarterly growth reflects only three months of activity and may understate the underlying momentum of the economy. An annual cumulative growth provides a clearer sense of the growth pace were it to persist over a full year.

For each quarter $t \in \{1895\text{Q}1, 1895\text{Q}2, \dots, 2016\text{Q}3, 2016\text{Q}4\}$, we use historical data from the quarters $\{1893\text{Q}1, 1893\text{Q}2, \dots, t-1\}$ as the training set. Both the `Old' and `New' methods introduced in this paper are implemented using the same code as in the simulation studies in Section~\ref{sec:simulation}.
Our proposed threshold-selection method in Section~\ref{sec:rule_of_thumb} yields varying values of $y_{\min}$ and $y_{\max}$ across quarters.
The median values across the quarters used in our out-of-sample prediction exercises are 2.87 for $y_{\min}$ and $-1.73$ for $y_{\max}$.

\begin{figure}[tbp]
    \centering
    \includegraphics[width=0.85\textwidth]{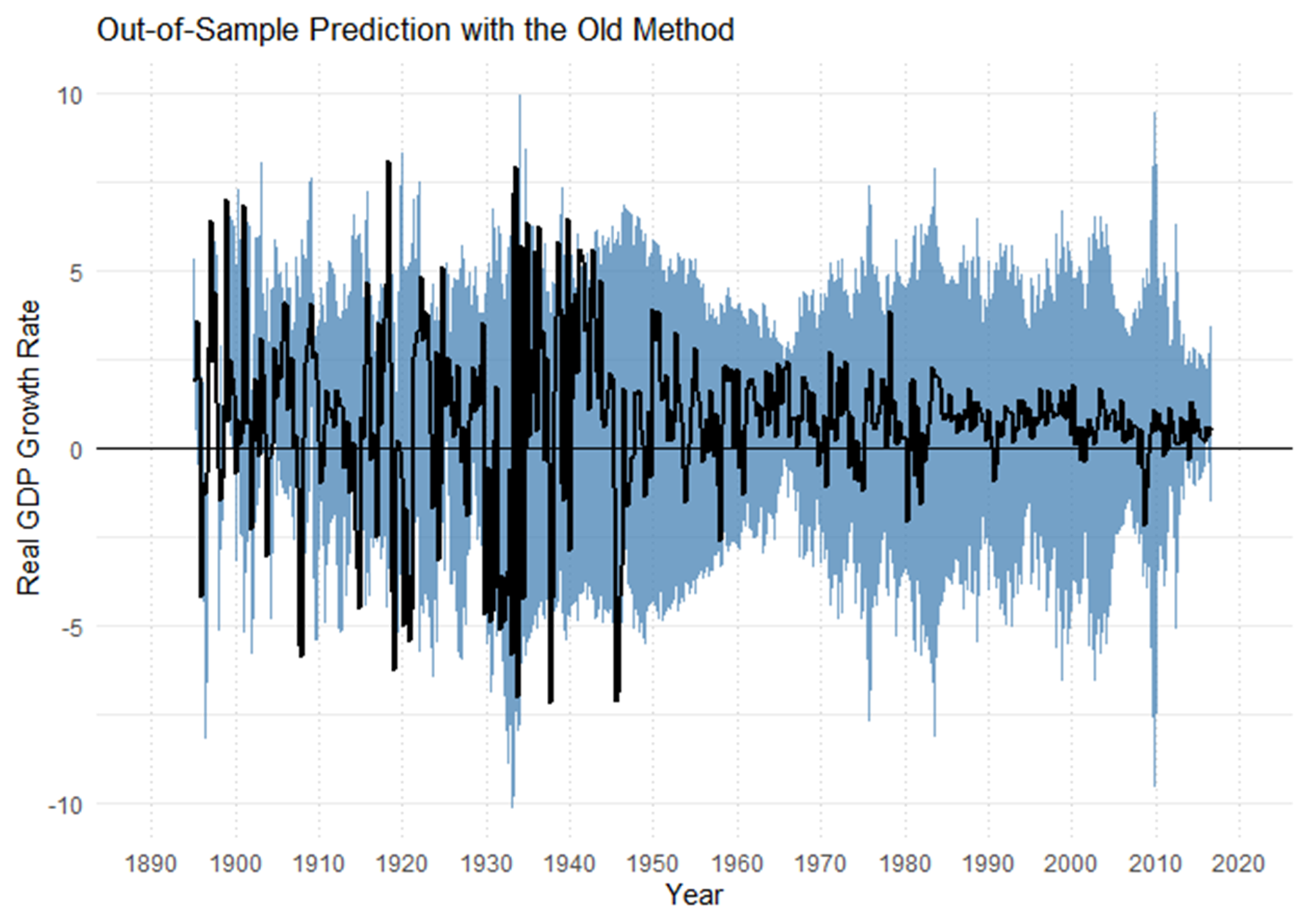}\\${}$\\
    \includegraphics[width=0.85\textwidth]{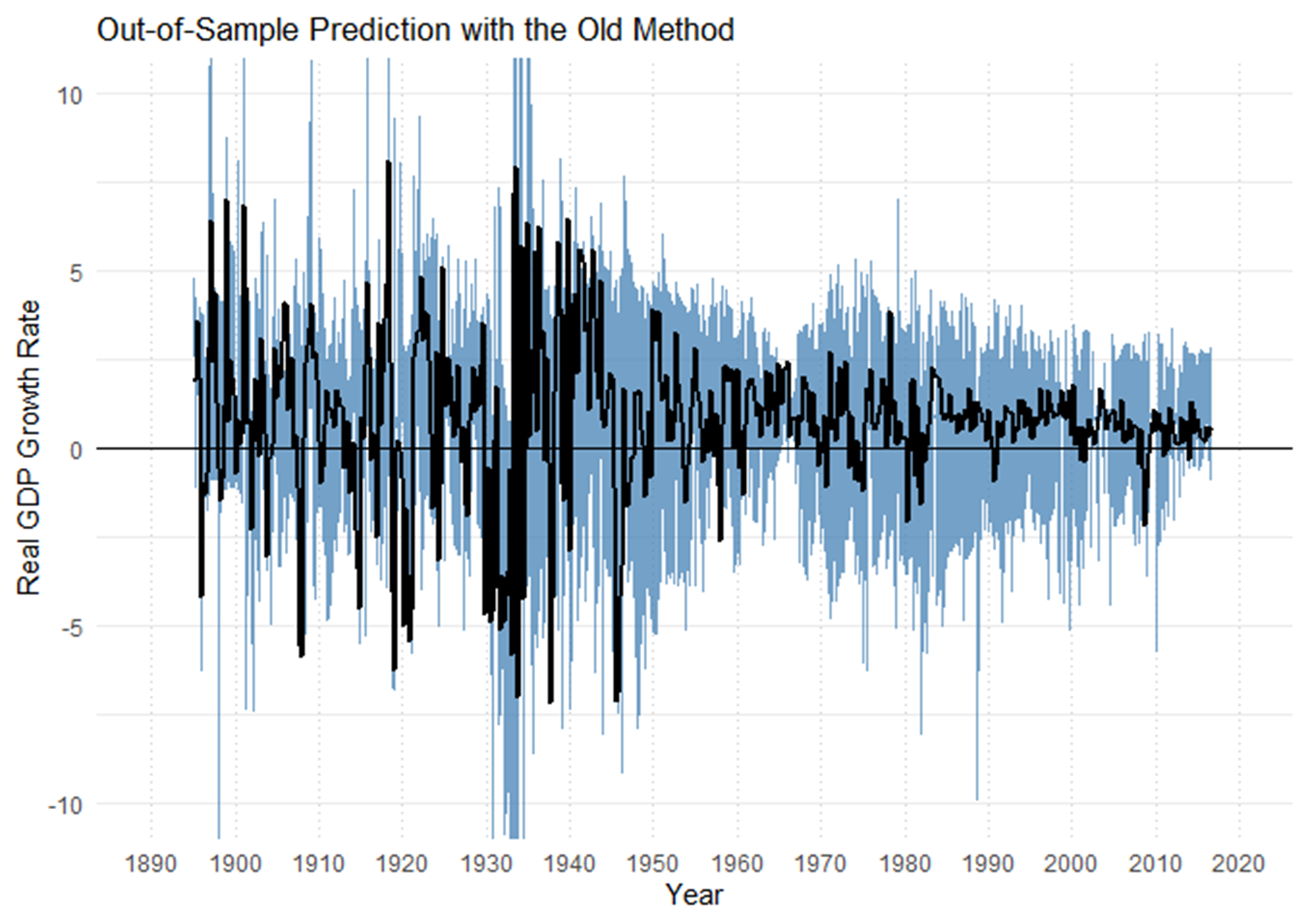}
    \caption{Out-of-sample predictions of the 5th and 95th percentiles of the real GDP growth rate between 1895Q1 and 2016Q4 based on the `Old' method (top) and the `New' method (bottom), with the prediction ranges indicated by the shares. The black lines indicate the actual real GDP growth rates.}
    \label{fig:app:pred}
\end{figure}

Figure \ref{fig:app:pred} illustrates the out-of-sample predictions. 
The top panel presents results obtained using the `Old' method, while the bottom panel displays results based on the `New' method. In each panel, the shaded area represents the range between the predicted 5th and 95th percentiles of real GDP growth rates four quarters ahead, and the solid black line indicates the realized real GDP growth rates.

Observe that the realized GDP growth rates exhibit considerable volatility up to the end of World War~II in 1945, whereas they become markedly more stable thereafter. Accordingly, we expect a valid prediction method to capture this pattern by producing wider predicted intervals before 1945 and narrower intervals in the postwar period.

Consider the prediction intervals generated by the `Old' method in the top panel, which display similar magnitudes of volatility before and after 1945. This pattern suggests that the intervals are excessively wide in the post-1945 period. In particular, after 1970, the actual GDP growth rates rarely, if ever, fall outside the predicted range. This is problematic, as predictions based on the 5th and 95th percentiles are expected to exclude approximately 10 percent of the actual realizations.

In contrast, the predictions generated by the `New' method in the bottom panel exhibit narrower intervals after 1945 than before, appropriately adjusting to the reduced volatility of actual growth rates in the postwar period.

As mentioned above, the prediction intervals between the 5th and 95th percentiles should ideally capture approximately 90 percent of the realized real GDP growth rate observations. This implies that, over the 488-quarter out-of-sample prediction horizon, the realized growth rates should fall below the 5th percentile in about 24 quarters and exceed the 95th percentile in another 24 quarters.

Column group (I) of Table~\ref{tab:frequency} reports that, over the 488 quarters, the `Old' method yields realized growth rates that exceed the 95th percentile only 16 times and fall below the 5th percentile just 11 times. These correspond to frequencies of 3.3\% and 2.3\%, respectively, compared to the target probability of 5.0\% for each tail.

\begin{table}[tbp]
\centering
\begin{tabular}{llllll}
\hline\hline
& \multicolumn{2}{c}{(I) `Old' Method} && \multicolumn{2}{c}{(II) `New' Method}\\
\cline{2-3}\cline{5-6}
Event&
$Y_{t+1}$ was below&
$Y_{t+1}$ was above&&
$Y_{t+1}$ was below&
$Y_{t+1}$ was above\\
&
$\hat{Q}^{\text{Lower}}_{Y_{t+1}|X_t}(0.05|x_0)$ &
$\hat{Q}^{\text{Upper}}_{Y_{t+1}|X_t}(0.95|x_0)$
&&
$\hat{Q}^{\text{Lower}}_{Y_{t+1}|X_t}(0.05|x_0)$ &
$\hat{Q}^{\text{Upper}}_{Y_{t+1}|X_t}(0.95|x_0)$\\ 
\hline
Observed\\
Frequency & \multicolumn{1}{c}{3.3\% (16/488)} & \multicolumn{1}{c}{2.3\% (11/488)} && \multicolumn{1}{c}{5.3\% (26/488)} & \multicolumn{1}{c}{4.5\% (22/488)}\\
\\
Intended \\
Frequency & \multicolumn{1}{c}{5.0\%} & \multicolumn{1}{c}{5.0\%} && \multicolumn{1}{c}{5.0\%} & \multicolumn{1}{c}{5.0\%}\\
\hline\hline
\end{tabular}
\caption{Frequencies of the events that the realized real GDP growth rates were below the out-of-sample 5th predicted percentiles and above the 95th predicted percentiles between 1950 and 2020 based on each of the (I) `Old' and (II) `New' methods.}${}$\\
\label{tab:frequency}
\end{table}

In contrast, column group (II) shows that the `New' method yields nearly the targeted number of quarters in which the realized growth rate exceeds the 95th predicted percentile (26 times, or 5.3\%) and falls below the 5th predicted percentile (22 times, or 4.5\%), closely aligning with the target probability of 5.0\% for each tail.
Since year-ahead forecasts $(h = 4)$ overlap in quarterly data, the sequence of forecast errors is serially correlated. While this serial dependence would matter for formal hypothesis testing about predictive accuracy, it does not affect the raw coverage proportions reported in Table 1, which are purely descriptive counts of exceedances. Thus, the interpretation of Table 1 remains unaffected.

Finally, we use the full time series from 1893 to 2016 to generate out-of-sample predictions for future GaR. Specifically, we conduct a prediction exercise under the assumption that the current year exhibits the median values observed in the dataset. We evaluate GaR for the following year under two scenarios for the current-year FCI: (i) a ``normal year'' scenario, in which the FCI takes its historical median value; and (ii) a ``crisis year'' scenario, in which the FCI is set to the level observed during the 2008Q4 financial crisis. For each scenario, we apply both the `Old' and `New' methods to estimate the tail densities of the predicted real GDP growth distribution for the following year.

\begin{figure}[tbp]
    \centering
    \includegraphics[width=0.7\textwidth]{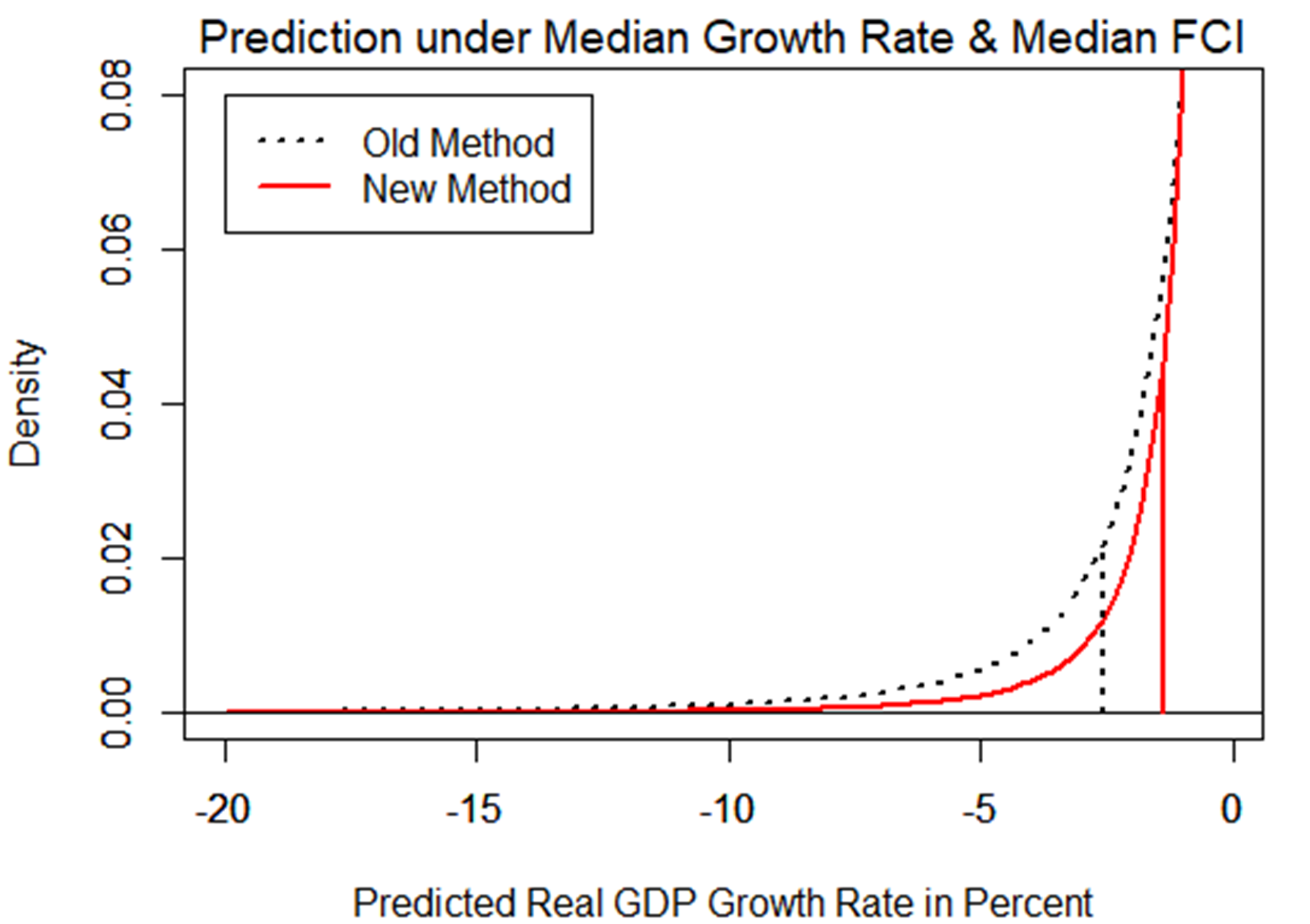}
    \bigskip\\
    \includegraphics[width=0.7\textwidth]{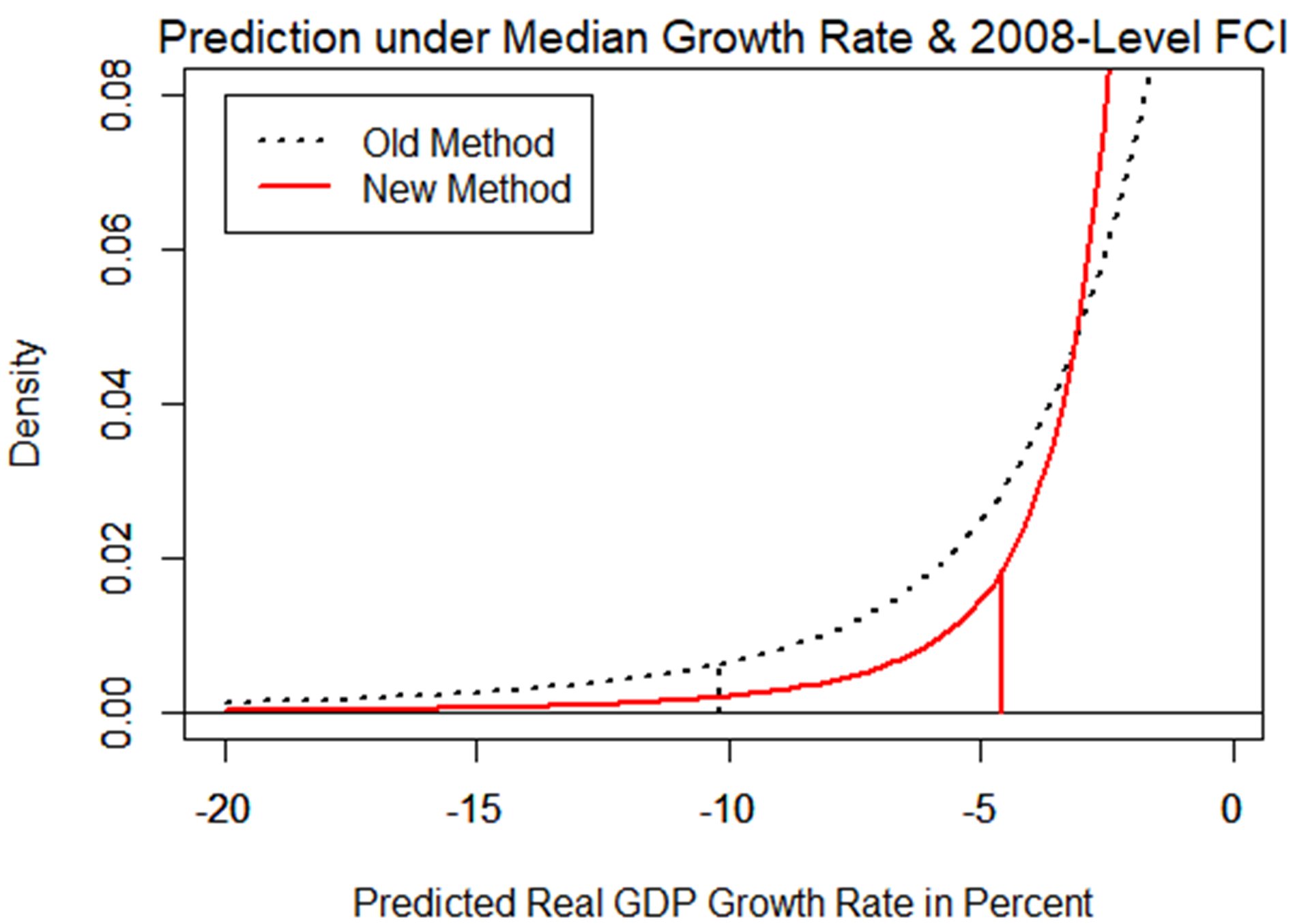}
    \caption{Lower tails of the PDFs of predicted real GDP growth rates, conditional on a current-year real GDP growth rate equal to the historical median. The top panel presents the distribution under the ``normal year'' scenario, where the current-year Financial Conditions Index (FCI) equals its historical median. The bottom panel presents the distribution under the ``crisis year'' scenario, where the FCI equals its 2008Q4 level. In each panel, the dotted line corresponds to predictions from the `Old' method, and the solid line corresponds to the `New' method. Vertical line segments indicate the 5th percentiles of the predicted real GDP growth distributions. All predictions are based on the full time series from 1880 to 2020.}${}$\\
    \label{fig:app:density}
\end{figure}

Figure \ref{fig:app:density} presents these tail density plots.
The top panel presents the predicted distribution under the ``normal year'' scenario, where the current-year Financial Conditions Index (FCI) equals its historical median. The bottom panel shows the distribution under the ``crisis year'' scenario, where the FCI is set to its 2008Q4 level. In each panel, the dotted line corresponds to predictions from the `Old' method, and the solid line represents those from the `New' method. Vertical line segments indicate the 5th percentiles of the predicted real GDP growth distributions.

First, consider the predictions based on the median FCI, as shown in the top panel. In this case, the `Old' and `New' methods yield predicted fifth percentile growth rates of $-$2.6 and $-$1.4, respectively. While the `Old' method produces a wider interval, as documented above, the difference between the two methods is relatively modest in this scenario.

Second, consider the bottom panel, which presents predictions under the 2008-level FCI. Here, the discrepancy between the `Old' and `New' methods becomes substantial. Specifically, the `Old' method predicts a fifth percentile growth rate of $-10.2$, whereas the `New' method yields a significantly milder value of $-4.6$. Given that the 5th predicted percentile is expected to materialize roughly once in five years (i.e., 5 out of 100 quarters) and the smallest observed U.S.\ GDP growth rate in the 491 quarters of available history is only $-7.2\%$, the predicted value of $-10.2$ produced by the `Old' method appears too low to be plausible given the empirical benchmarks.

In light of these observations, let us set aside the counterfactual prediction analyses and return to the out-of-sample prediction exercises illustrated in Figure~\ref{fig:app:pred}. 
The year-ahead prediction produced by the `Old' method based on 2008Q4 yields not only an implausibly low predicted 5th percentile but also an implausibly high predicted 95th percentile. 
In other words, the `Old' method predicts explosive volatility rather than a downturn. 
By contrast, the `New' method produces a moderately low predicted 5th percentile without inducing a jump in the predicted 95th percentile, indicating that it successfully captures the asymmetric nature of the downturn, as opposed to a spike in volatility, unlike the `Old' method.

While we focused on the episode of the 2008 financial crisis because it is more immediate in our memory, similar patterns are observed in other financial crises in U.S.\ history. 
For instance, the year-ahead predicted 5th and 95th percentiles produced by the `Old' method after the Wall Street Crash of 1929Q4 are \(-6.4\) and \(6.8\), respectively. 
By contrast, our `New' method predicts the 5th and 95th percentiles to be \(-2.5\) and \(0.9\), respectively---again offering more plausible downturn predictions rather than the implausibly large volatility spikes implied by the `Old' method.
Although the quantitative magnitudes differ, a similar observation can be made for other crashes, such as the Black Monday stock market crash.

\section{Summary and Discussions}\label{sec:summary}

This paper discusses certain limitations of existing econometric methods concerning Growth-at-Risk (GaR) and, in light of these limitations, proposes a novel method that is robust to them. We demonstrate the efficacy of our proposed method, both theoretically and through numerical simulations, in comparison to existing alternatives.

With this new tool, we study the GaR in the long history of the U.S.
Applying our novel method to a long-term U.S. macro-financial dataset,
we find that our approach successfully predicts low but plausible GaR values that capture the impact of the financial crisis, whereas the existing alternative predicts implausibly low GaR values. Moreover, our method achieves the intended frequencies of GaR, while the existing method yields overly extreme predictions. In light of these findings, we recommend adopting our proposed method as a more robust alternative.

While our primary focus is on the GaR framework, the proposed method is broadly applicable to other contexts involving the study of extreme outcomes, including
Financial Stability at Risk \citep{adrian2016covar},
Employment at Risk \citep{ramey2018government},
Growth at Risk \citep{adrian2019vulnerable},
Inflation at Risk \citep{loria2022inflation},
and Term Structure of Growth at Risk \citep{adrian2022term},
among others.
We hope this paper will inspire further empirical research on these and other topics in economics and finance. 

Finally, we conclude this paper by discussing several promising directions for future theoretical research. 
First, although our specification~\eqref{eq:v_exp} for the tail exponent follows the existing literature and our simulation studies demonstrate its robustness (see Section \ref{sec:simulations_comment_additional} and Appendix \ref{sec:additional_simulation:misspecification}), it may nevertheless be misspecified. 
To address this concern, one could consider a semiparametric model of the form 
\( v(x) = \exp(g(x'\beta)) \), where \( g \) is a nonparametric function. 
Such a generalization may offer additional robustness.
Second, our Assumption~P requires strict stationarity of \(\{Y_t, X_t\}\), whereas the empirical data used in Section~\ref{sec:history} appear to exhibit structural breaks. To make our method more consistent with the empirical setting, it would be useful to incorporate regime switching into our robust GaR model. Selecting structural break points and conducting post-selection estimation would provide a more robust framework.
Third, while we focus on U.S.\ macro-historical data in this paper, it would be worthwhile to extend the empirical analyses to other countries and datasets. 
For example, a promising example is to explore the relationship between the tail risk of currency returns and financial conditions such as implied volatility of stocks, bonds, and currencies.
Fourth, our method is designed primarily to deliver robust estimates of the extreme tails, which are the core objects in GaR analysis. To produce a full conditional density, one may splice our Pareto-type tails onto any baseline estimator for the interior quantiles (e.g., interpolated quantile regressions or kernel methods), and renormalize the resulting distribution so that it integrates to one. This approach retains full tail flexibility while ensuring global coherence. 
Alternatively, our estimated tail indices can be embedded as constraints in flexible parametric or semi-parametric density families.
We leave these extensions as important directions for future research.

\newpage
\appendix
\section*{Appendix}
\sloppy
\section{Mathematical Proofs}

Before proceeding with proofs, we collect some important definitions to be used in the proofs.

First, we say that the conditional distribution of $Y_{t+h}$ given $X_t=x$ is within the domain of attraction (DoA) of some extreme value distribution if, for i.i.d. draws $Z_{1},...,Z_{m}$ from $F_{Y_{t+h}|X_t}\left( \ \cdot \ |x\right) 
$, there exist sequences of constants $a_{m}$ and $b_{m}$ such that as $m\rightarrow\infty$
\begin{equation*}
\frac{\max_{m}\{Z_{1},...,Z_{m}\}-b_{m}}{a_{m}}\overset{d}{\rightarrow }V,
\end{equation*}
where $V$ follows the generalized extreme value distribution whose CDF is given by
\begin{equation*}
G_{\xi \left( x\right) }\left( v\right) =\exp \left( -\left( 1+\xi \left(x\right) v\right) ^{-1/\xi \left( x\right) }\right) .
\end{equation*}
This DoA condition holds for a broad class of nonparametric distributions, including the skewed-t in particular.
Thus, \citet{adrian2019vulnerable} implicitly assume this DoA condition too.

Second, we say that the conditional distribution of $Y_{t+h}$ given $X_t=x$ satisfies the regularly varying (RV) tail condition if
\begin{equation}
1-F_{Y|X=x}(y)\sim C_{1}y^{-v(x)}\text{ as }y\rightarrow \infty,  \label{eq:RV}
\end{equation}
where $v(x)$ denotes the Pareto exponent.
In this case, it holds that as $y_{\min}\rightarrow\infty$
\begin{equation*}
\mathbb{E}\left[ Y|Y>y_{\min },X=x\right] \sim y_{\min }\frac{v(x)}{v(x)-1}
\end{equation*}
This RV condition holds for a broad class of nonparametric distributions, including the skewed-t in particular.
Thus, \citet{adrian2019vulnerable} implicitly assume this RV condition too.

\subsection{Proof of Proposition \ref{prop:imposs}}\label{sec:prop:imposs}

\begin{proof}
For simplicity of writing, we abbreviate the notations $Y_{t+h}$ and $X_t$ as $Y$ and $X$, respectively, in this proof. 
We focus on the right tail, as the argument for the left tail follows symmetrically. 
Our proof strategy is outlined as follows:
on the one hand, the skewed-t distribution implies the ratio of two conditional quantiles varies with $x$ through the exponent $v(x)$;
on the other hand, a linear quantile regression structure implies that such a ratio remains constant over $x$, leading to the conclusion that $v(x)=\bar{v}$ for some constant $\bar{v}$.

We first investigate the DoA assumption.
Specifically, since the skewed-t distribution satisfies the DoA assumption with $\xi \left(x\right) =1/v\left( x\right)>0 $, it follows that $Q_{Y|X}\left( \tau |x\right) \sim c_1(x)(1-\tau)^{-1/v(x)} \rightarrow \infty $ as $\tau \rightarrow 1 $. Therefore, we obtain that
\begin{equation}\label{eq:Q_ratio}
    \frac{Q_{Y|X}\left( 1-\frac{1}{t\tau }|x\right) }{Q_{Y|X}\left( 1-\frac{1}{t}|x\right) }\sim \tau^{1/v(x)}
\end{equation}
for any $\tau>0$ as $t\to\infty$. Such a ratio changes over $x$ if and only if $v(x)$ is non-constant.

Now we explore the linear quantile regression structure.
Let $x=(1,\dot{x}')'$ so that $Q_{Y|X}\left( \tau |x\right) = \alpha(\tau) + \dot{x}'\beta(\tau)$. 
Let $j^*$ denote the coordinate of $\beta$ that has the heaviest tail,  
i.e., $\lim_{\tau\to1}\beta_j(\tau)/\beta_{j^*}(\tau) = 0$ for all $j\neq j^*$.\footnote{If two components have the same order, i.e., $\lim_{\tau\to1}\beta_j(\tau)/\beta_{j^*}(\tau)=c$ for some constant $c\in(0,\infty)$ and some $j$, then combine components $j$ and $j^*$ and consider $x_{j^*}+cx_{j}$ to be the dominating component.} 
Now let $h=\lim_{\tau\to1}\beta_{j^*}(\tau)/\alpha(\tau)\ge 0$.
We now proceed with each of the three cases, $h=0$, $h=\infty$, and $h\in(0,\infty)$, separately. 

\bigskip\noindent
\underline{Case 1: $h=0$.} 
In this case, $\beta_{j^*}(1-1/t)=o(\alpha(1-1/t))$ as $t\rightarrow \infty$ and hence
\begin{align*}
\lim_{t\to\infty}\frac{Q_{Y|X}\left( 1-\frac{1}{t\tau }|x\right) }{Q_{Y|X}\left( 1-\frac{1}{t}|x\right) } 
= \lim_{t\to\infty}\frac{\alpha\left( 1-\frac{1}{t\tau }\right) +\dot{x}_{j^*}\beta_{j^*}(1-\frac{1}{t\tau})}{\alpha\left( 1-\frac{1}{t}\right)+\dot{x}_{j^*}\beta_{j^*}(1-\frac{1}{t}) }
= \lim_{t\to\infty}\frac{\alpha\left( 1-\frac{1}{t\tau }\right) }{\alpha\left( 1-\frac{1}{t}\right) },
\end{align*}
which does not depend on $\dot{x}$.

\bigskip\noindent
\underline{Case 2: $h=\infty$.} 
In this case, $\alpha(1-1/t)=o(\beta_{j^*}(1-1/t))$ as $t\rightarrow \infty$ and hence
\begin{align*}
\lim_{t\to\infty}\frac{Q_{Y|X}\left( 1-\frac{1}{t\tau }|x\right) }{Q_{Y|X}\left( 1-\frac{1}{t}|x\right) } 
= \lim_{t\to\infty}\frac{\alpha\left( 1-\frac{1}{t\tau }\right) +\dot{x}_{j^*}\beta_{j^*}(1-\frac{1}{t\tau})}{\alpha\left( 1-\frac{1}{t}\right)+\dot{x}_{j^*}\beta_{j^*}(1-\frac{1}{t}) }
= \lim_{t\to\infty}\frac{\beta_{j^*}\left( 1-\frac{1}{t\tau }\right) }{\beta_{j^*}\left( 1-\frac{1}{t}\right) },
\end{align*}
which does not depend on $\dot{x}$.

\bigskip\noindent
\underline{Case 3: $h\in(0,\infty)$.} 
In this case, $h\alpha(1-1/t)\sim \beta_{j^*}(1-1/t)$ as $t\rightarrow \infty$ and hence
\begin{align*}
\lim_{t\to\infty}\frac{Q_{Y|X}\left( 1-\frac{1}{t\tau }|x\right) }{Q_{Y|X}\left( 1-\frac{1}{t}|x\right) } 
= \lim_{t\to\infty}\frac{(\dot{x}_{j^*}+h^{-1})\beta_{j^*}\left( 1-\frac{1}{t\tau }\right) }{(\dot{x}_{j^*}+h^{-1})\beta_{j^*}\left( 1-\frac{1}{t}\right)}
= \lim_{t\to\infty}\frac{\beta_{j^*}\left( 1-\frac{1}{t\tau }\right) }{\beta_{j^*}\left( 1-\frac{1}{t}\right)},
\end{align*}
which does not depend on $\dot{x}$.

Combining three cases yields that $\lim_{t\to\infty}\frac{Q_{Y|X}\left( 1-\frac{1}{t\tau }|x\right) }{Q_{Y|X}\left( 1-\frac{1}{t}|x\right) }$ does not depend on $\dot{x}$.
However, recall \eqref{eq:Q_ratio} that
\begin{align*}
\frac{Q_{Y|X}\left( 1-\frac{1}{t\tau }|x\right) }{Q_{Y|X}\left( 1-\frac{1}{t}|x\right) }\rightarrow \tau ^{1/v\left( x\right) }
\end{align*}
as $t\rightarrow \infty$.
Therefore, it follows that $v\left(x\right) =\bar{v}$ for some constant $\bar{v}$, which does not depend on $x$.
\end{proof}

\subsection{Proof of Proposition \ref{prop:tce}}\label{sec:prop:tce}

\begin{proof}
For simplicity of writing, we abbreviate the notations $Y_{t+h}$ and $X_t$ as $Y$ and $X$, respectively.
We introduce the tail conditional expectation (TCE) defined by
\begin{equation*}
T_{Y|X}\left( \tau |x\right) =\mathbb{E}\left[ Y|Y>Q_{Y|X}\left( \tau |x\right) ,X=x\right] .
\end{equation*}
Since the skewed-t conditional distribution with $v(x)$ degrees of freedom satisfies the RV condition \eqref{eq:RV} with $v(x)$ as the tail exponent,
we have
\begin{equation*}
f_{Y|X=x}(y)\sim C_{1}v(x)y^{-v(x)-1},
\end{equation*}
where $A\sim B$ means $A/B\rightarrow 1$ as $y\rightarrow \infty $. 
It therefore follows that
\begin{align}
\text{ \ \ \ }\frac{T_{Y|X}\left( \tau |x\right) }{Q_{Y|X}\left( \tau |x\right) }
& \overset{(i)}{\sim }\frac{C_{1}v(x)\int_{Q_{Y|X}\left( \tau |x\right)}^{\infty }y^{-v(x)}dy}{\left( 1-\tau \right) Q_{Y|X}\left( \tau |x\right) } \notag\\
& \overset{(ii)}{=}C_{1}v(x)Q_{Y|X}\left( \tau |x\right) ^{-v(x)}\int_{1}^{\infty }t^{-v(x)}dt \notag\\
& \overset{(iii)}{\sim }\int_{1}^{\infty }t^{-v(x)}dt \notag\\
& = \left\{ 
\begin{array}{cl}
\frac{v(x)}{v(x)-1} & \text{if }v(x)>1 \\ 
\infty  & \text{otherwise,}
\end{array}
\right. 
\label{eq:t_over_q}
\end{align}
where 
step (i) is due to the RV condition \eqref{eq:RV}, 
step (ii) is by the change of variables from $y$ to $tQ_{Y|X}\left( \tau |x\right) $, and 
step (iii) follows from the fact that 
$$Q_{Y|X}\left( \tau |x\right) \sim \left[ C_{1}^{-1}(1-\tau )\right]^{-1/v(x)}$$ 
as $\tau \rightarrow 1$, which is further implied by the RV condition. 

Now, let $\tau(\pi) = 1-\pi$.
Then, observe that 
\begin{align}
LR_{Y|X}(\pi|x) 
=&
T_{Y|X}(\tau(\pi)|x)
\label{eq:lr_over_pi}
\end{align}
Since $\tau(\pi) \rightarrow 1$ as $\pi \rightarrow 0$, it follows from \eqref{eq:t_over_q}--\eqref{eq:lr_over_pi} that
\begin{align*}
\frac{LR_{Y|X}(\pi|x)}{Q_{Y|X}(\tau(\pi)|x)} 
\rightarrow
\left\{ 
\begin{array}{cl}
\frac{v(x)}{v(x)-1} & \text{if }v(x)>1 \\ 
\infty  & \text{otherwise}
\end{array}
\right.
\end{align*}
as $\pi \rightarrow 0$.
\end{proof}

\subsection{Proof of Theorem \ref{thm:asym}}\label{sec:thm:asym}

\begin{proof}
We focus on the proof for the extreme quantile. 
The proof for the longrise is very similar and omitted. 
For simplicity of writing, we abbreviate the notations $Y_{t+h}$ and $X_t$ as $Y$ and $X$, respectively.

Define $U\left( t|x_{0}\right) =F_{Y|X}^{-1}\left( 1-1/t|x_{0}\right) $. 
\citet[][Remark 3.2.7]{dehaan2006extreme} and our Assumption P imply
\begin{equation*}
\frac{U(ty|x_{0})}{U(t|x_{0})}\sim y^{1/v(x_{0})}\left( 1+t^{-\rho(x_{0})}\frac{y^{-\rho (x_{0})}-1}{-\rho (x_{0})}\right) .
\end{equation*}

Let $t_{\min }$ satisfy $U(t_{\min }|x_{0})=y_{\min }$. 
Note that
\begin{equation*}
1-\frac{1}{t_{\min }}=F_{Y|X}\left( y_{\min }|x_0\right) \Leftrightarrow t_{\min }=\frac{1}{1-F_{Y|X}\left( y_{\min }|x_0\right) }\equiv p_{\min }^{-1}.
\end{equation*}
Recall $1-\tau =\left( 1-\tau _{\min }\right) p_{\min }$. This yields
\begin{align}
Q_{Y|X}\left( \tau |x_{0}\right) =&U\left( t_{\min }y|x_{0}\right)  \notag\\
\sim &y_{\min }\left( 1-\tau _{\min }\right) ^{-1/v(x_{0})}\left(1+t_{\min }^{-\rho (x_{0})}\frac{y^{-\rho (x_{0})}-1}{-\rho (x_{0})}\right)
\label{eq:approx:q}
\end{align}
as $\tau \rightarrow 1$ (and $y_{\min }\rightarrow \infty $).

Now, we introduce the estimator $\hat{F}_{Y|X}\left( y_{\min }|x_{0}\right) $ and the corresponding $\hat{\tau}_{\min }$ such that 
\begin{equation*}
1-\tau =\left( 1-\hat{\tau}_{\min }\right) \left( 1-\hat{F}_{Y|X}\left(y_{\min }|x_{0}\right) \right) .
\end{equation*}
Observe that 
\begin{align}
\left( \frac{1-\hat{\tau}_{\min }}{1-\tau _{\min }}\right) ^{-1/v(x_{0})} =&\left( \frac{1-F_{Y|X}\left( y_{\min }|x_0\right) }{1-\hat{F}_{Y|X}\left( y_{\min }|x_0\right) }\right) ^{-1/v(x_{0})} \notag\\
\sim &\left( 1+\frac{\hat{F}_{Y|X}\left( y_{\min }|x_0\right) -F_{Y|X}\left(y_{\min }|x_0\right) }{1-\hat{F}_{Y|X}\left( y_{\min }|x_0\right) }\right)^{-1/v(x_{0})} \notag\\
\sim &1-\frac{1}{v\left( x_0\right) }\frac{\hat{F}_{Y|X}\left( y_{\min}|x_0\right) -F_{Y|X}\left( y_{\min }|x_0\right) }{1-\hat{F}_{Y|X}\left( y_{\min}|x_0\right) }
\label{eq:tau_ratio}
\end{align}
holds, provided that 
$$
\frac{\hat{F}_{Y|X}\left( y_{\min }|x_0\right) -F_{Y|X}\left( y_{\min }|x_0\right) }{1-\hat{F}_{Y|X}\left( y_{\min }|x_0\right) }=o_{p}(1).
$$
This indeed holds under Assumptions P and R.(i), as 
\begin{align*}
&\frac{\hat{F}_{Y|X}\left( y_{\min }|x_0\right) -F_{Y|X}\left( y_{\min}|x_0\right) }{1-\hat{F}_{Y|X}\left( y_{\min }|x_0\right) } \\
=&\frac{\left( T^{1/2}r_{T}\right) ^{-1}\left( 1-F_{Y|X}\left( y_{\min}|x_{0}\right) \right) ^{1/2}}{\left( 1-F_{Y|X}\left( y_{\min }|x_{0}\right)\right) +\left( T^{1/2}r_{T}\right) ^{-1}\left( 1-F_{Y|X}\left( y_{\min}|x_{0}\right) \right) ^{1/2}} \\
=&O_{p}\left( \left( 1-F_{Y|X}\left( y_{\min }|x_{0}\right) \right)^{-1/2}\left( T^{1/2}r_{T}\right) ^{-1}\right)  \\
=&O_{p}\left( y_{\min }^{v(x_{0})/2}\left( T^{1/2}r_{T}\right)^{-1}\right)  \\
=&o_{p}(1).
\end{align*}

Substituting \eqref{eq:tau_ratio} this into \eqref{eq:approx:q}, we obtain
\begin{align*}
&\frac{\hat{Q}^{Upper}_{Y|Y>y_{\min },X}(\hat{\tau}_{\min }|x_{0})}{Q_{Y|Y>y_{\min},X}(\tau _{\min }|x_{0})} \\
=&\frac{\hat{Q}_{Y|Y>y_{\min },X}(\hat{\tau}_{\min }|x_{0})}{Q_{Y|Y>y_{\min},X}(\hat{\tau}_{\min }|x_{0})}\frac{Q_{Y|Y>y_{\min },X}(\hat{\tau}_{\min}|x_{0})}{Q_{Y|Y>y_{\min },X}(\tau _{\min }|x_{0})} \\
=&\left( 1-\hat{\tau}_{\min }\right) ^{-1/\hat{v}(x_{0})+1/v(x_{0})}\left( 1+t_{\min }^{-\rho (x_{0})}\frac{\left( 1-\hat{\tau}_{\min}\right) ^{\rho (x_{0})}-1}{-\rho (x_{0})}\right)  \\
&\times \left( 1-\tau _{\min }\right) ^{-1/\hat{v}(x_{0})+1/v(x_{0})}\left( 1+t_{\min }^{-\rho (x_{0})}\frac{\left( 1-\tau _{\min}\right)^{\rho (x_{0})}-1}{-\rho (x_{0})}\right)  \\
\sim &\left[ 1+\log \left( 1-\hat{\tau}_{\min }\right) \exp \left(-x_{0}^{\prime }\beta \right) x_{0}^{\prime }\left( \hat{\beta}-\beta\right) +t_{\min }^{-\rho (x_{0})}\frac{\left( 1-\hat{\tau}_{\min }\right)^{\rho (x_{0})}-1}{-\rho (x_{0})}\right]  \\
&\times \left( \frac{1-\hat{\tau}_{\min }}{1-\tau_{\min }}\right)^{-1/v(x_{0})}\frac{\left( 1+t_{\min }^{-\rho (x_{0})}\frac{\left( 1-\hat{\tau}_{\min}\right) ^{\rho (x_{0})}-1}{-\rho (x_{0})}\right) }{\left(1+t_{\min }^{-\rho (x_{0})}\frac{\left( 1-\tau _{\min}\right) ^{\rho (x_{0})}-1}{-\rho (x_{0})}\right) } \\
\sim &1+\log \left( 1-\hat{\tau}_{\min }\right) \exp \left( -x_{0}^{\prime}\beta \right) x_{0}^{\prime }\left( \hat{\beta}-\beta \right)  \\
&-\frac{1}{v \left( x_{0}\right) }\left( \frac{\hat{F}_{Y|X}\left(y_{\min }|x_{0}\right) -F_{Y|X}\left( y_{\min }|x_{0}\right) }{1-\hat{F}_{Y|X}\left( y_{\min }|x_{0}\right) }\right) +O\left( t_{\min }^{-\rho(x_{0})}\right) .
\end{align*}
Therefore, we can write
\begin{align*}
&\sqrt{T}r_{T}\left( \frac{\hat{Q}^{Upper}_{Y|Y>y_{\min },X}(\hat{\tau}_{\min}|x_{0})}{Q_{Y|Y>y_{\min },X}(\tau _{\min }|x_{0})}-1\right)  \\
\sim &\sqrt{T}r_{T}\log \left( 1-\hat{\tau}_{\min }\right) \exp \left(-x_{0}^{\prime }\beta \right) x_{0}^{\prime }\left( \hat{\beta}-\beta\right)  \\
&+\frac{\sqrt{T}r_{T}}{v\left( x_{0}\right) }\left( \frac{\hat{F}_{Y|X}\left( y_{\min }|x_{0}\right) -F_{Y|X}\left( y_{\min }|x_{0}\right) }{1-\hat{F}_{Y|X}\left( y_{\min }|x_{0}\right) }\right) +O\left( \sqrt{T}r_{T}t_{\min }^{-\rho (x_{0})}\right)  \\
\equiv &A_{1T}+A_{2T}+A_{3T}.
\end{align*}%
The first item $A_{1T}$ is $O_{p}\left( \sqrt{T/T_0}r_{T}\log\left( 1-\tau \right) \right) $ given Assumption R.(ii) and the fact $T_{0}/T=O_{p}\left( \mathbb{E}\left[ y_{\min }^{-v(X_{t})}\right]\right) $ \citep[][Theorem 1]{wang2009tail}. 
The third item $A_{3T}$ is $O\left( \sqrt{T}r_{T}y_{\min }^{-v(x_{0})\rho (x_{0})}\right) =o(1)$ given Assumption R.(iii). The second term $A_{2T}$ satisfies
\begin{equation*}
\Sigma _{F}\left( y_{\min },y_{\min }|x_{0}\right) ^{-1/2}A_{2T} \overset{d}{\rightarrow} \mathcal{N}\left( 0,1\right) 
\end{equation*}%
as implied by Assumption N. The proof is complete by combining three items.
\end{proof}

\section{Additional Simulation Studies}\label{sec:additional_simulation}

This appendix section presents additional simulation results that were omitted from Section~\ref{sec:simulation} in the main text. 
Appendix~\ref{sec:additional_simulation:fixed} presents simulations in which fixed threshold parameters are used instead of data-driven choices. 
Appendix~\ref{sec:additional_simulation:constant} presents simulations based on a design where the tail exponent model is constant in $X_t$. 
Appendix~\ref{sec:additional_simulation:misspecification} presents simulations based on a design where the tail exponent model is misspecified.

\subsection{Fixed Threshold Parameters}\label{sec:additional_simulation:fixed}

The simulation studies presented in Section~\ref{sec:simulation} of the main text use data-driven choices of the threshold parameters $y_{\max}$ and $y_{\min}$. This appendix section examines how these data-driven choices compare with simpler alternatives. Specifically, fixed thresholds set at the 90th and 10th percentiles of the empirical distribution of $Y_t$ for $y_{\max}$ and $y_{\min}$, respectively.
The simulation setting is identical to that introduced in Section~\ref{sec:simulation}, except for the choice rules for $y_{\max}$ and $y_{\min}$. 

\begin{figure}[tbp]
    \centering
    \hspace{1cm}{\footnotesize Fixed Thresholds} \hspace{5cm} {\footnotesize Data-Driven Thresholds}\\
    \includegraphics[width=0.49\textwidth]{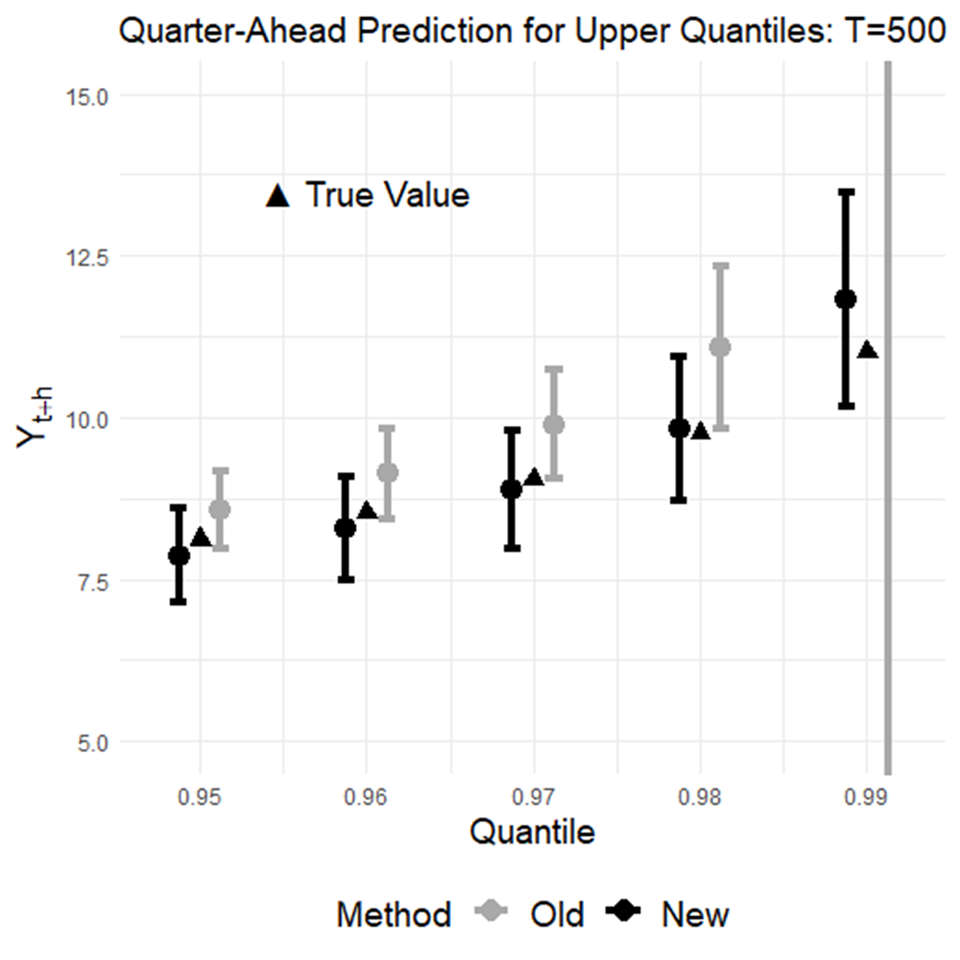}
    \includegraphics[width=0.49\textwidth]{fig_dd_quarter_upper_t0500.png}
    \\${}$\\
    \hspace{1cm}{\footnotesize Fixed Thresholds} \hspace{5cm} {\footnotesize Data-Driven Thresholds}\\
    \includegraphics[width=0.49\textwidth]{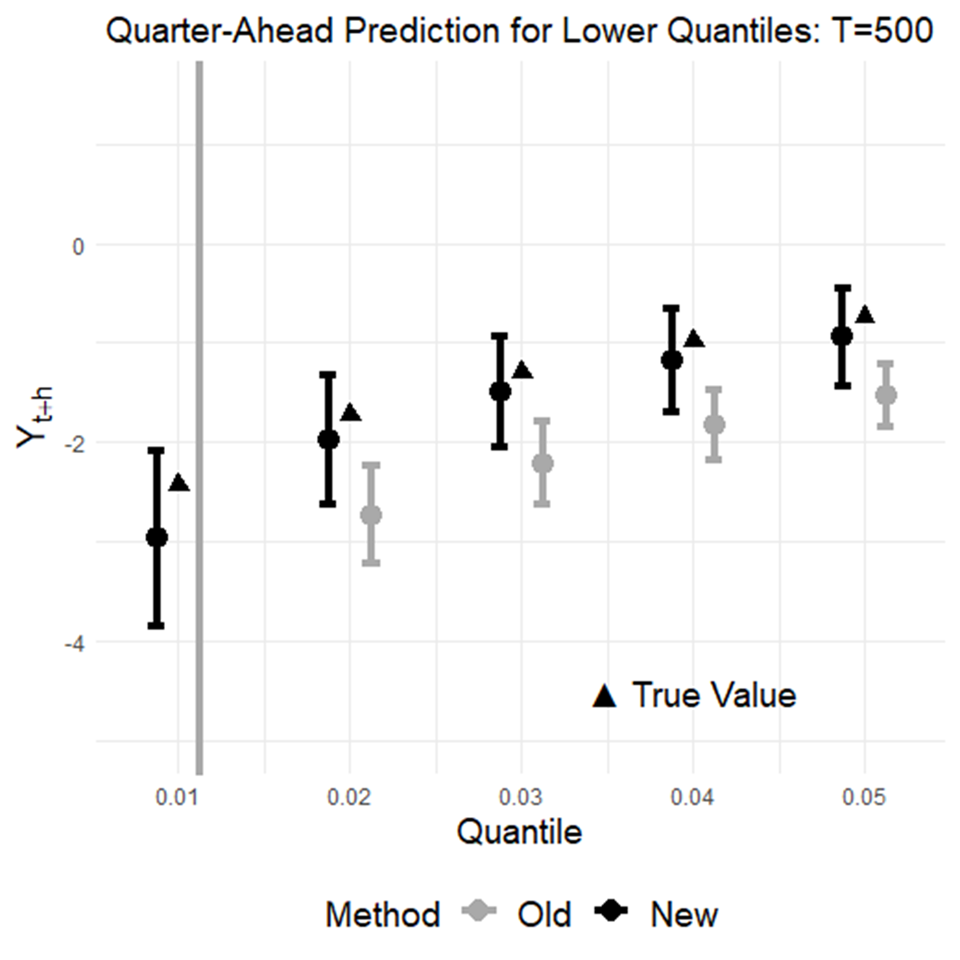}
    \includegraphics[width=0.49\textwidth]{fig_dd_quarter_lower_t0500.png}
    \caption{Simulation results comparing the performance of the proposed method (in black) and the existing method (in gray) for quarter-ahead predictions of growths. The predictions are conditional on the average values of $(X_{t1},X_{t2})= x_0 :=(2.732,0.007)$. The upper panels present results for the upper tail ($\tau \in \{0.95,0.96,0.97,098,0.99\}$), while the lower panels show results for the lower tail ($\tau \in \{0.01,0.02,0.03,0.04,0.05\}$). Dots represent simulation averages, bars represent the Gaussian interquartile ranges, and triangles denote the true values. The left column reports results based on fixed thresholds, $y_{\min}$ and $y_{\max}$, while the right column reports results based on data-driven choices of these thresholds.
    The results are based on 2,500 Monte Carlo iterations.}
    \label{fig:quarter_ahead_fixed}
\end{figure}

\begin{figure}[tbp]
    \centering
    \hspace{1cm}{\footnotesize Fixed Thresholds} \hspace{5cm} {\footnotesize Data-Driven Thresholds}\\
    \includegraphics[width=0.49\textwidth]{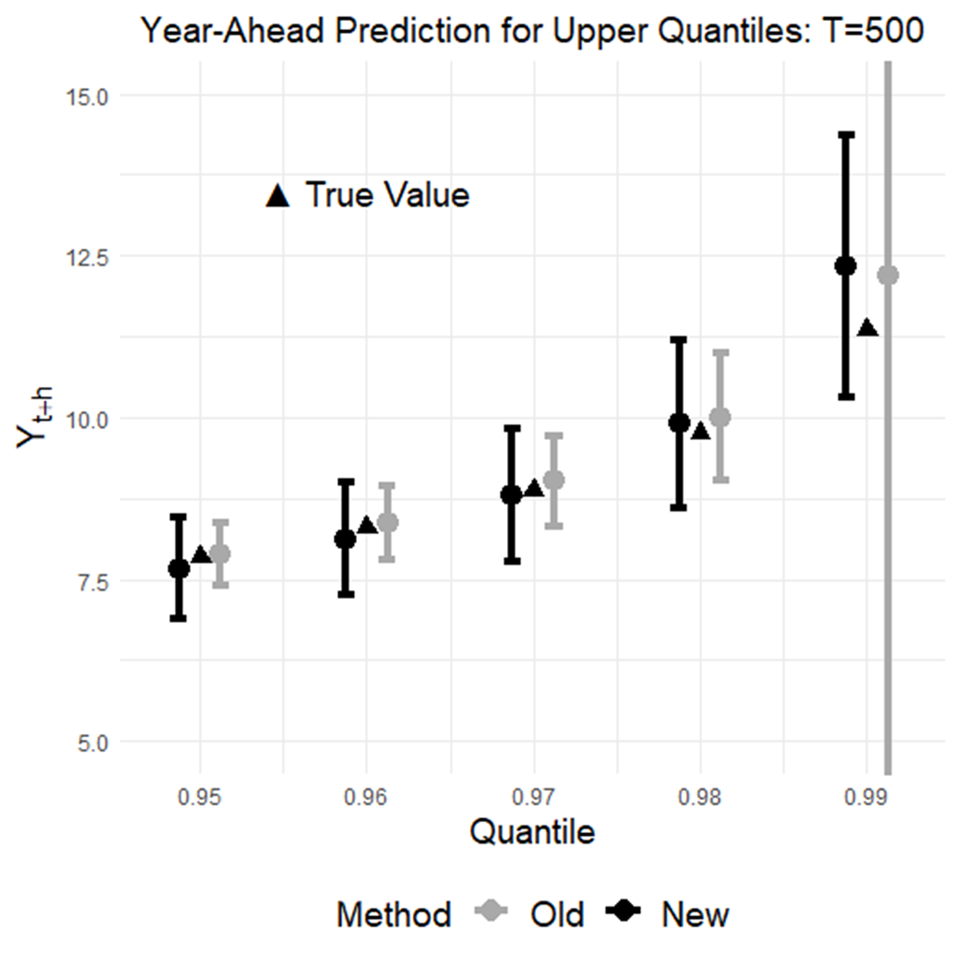}
    \includegraphics[width=0.49\textwidth]{fig_dd_year_upper_t0500.png}
    \\${}$\\
    \hspace{1cm}{\footnotesize Fixed Thresholds} \hspace{5cm} {\footnotesize Data-Driven Thresholds}\\
    \includegraphics[width=0.49\textwidth]{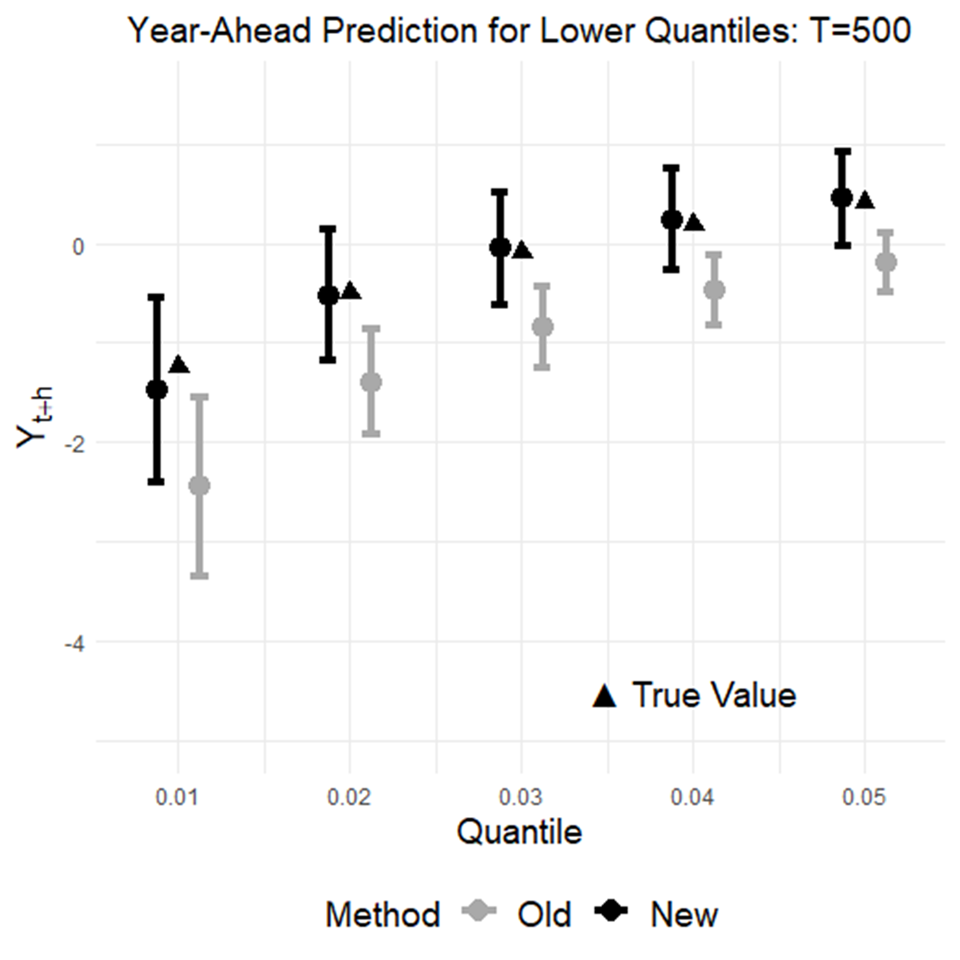}
    \includegraphics[width=0.49\textwidth]{fig_dd_year_lower_t0500.png}
    \caption{Simulation results comparing the performance of the proposed method (in black) and the existing method (in gray) for year-ahead predictions of growths. The predictions are conditional on the average values $(X_{t1},X_{t2})=x_0 :=(2.761,0.018)$. The upper panels present results for the upper tail ($\tau \in \{0.95,0.96,0.97,098,0.99\}$), while the lower panels show results for the lower tail ($\tau \in \{0.01,0.02,0.03,0.04,0.05\}$). Dots represent simulation averages, bars represent the Gaussian interquartile ranges, and triangles denote the true values. The left column reports results based on fixed thresholds, $y_{\min}$ and $y_{\max}$, while the right column reports results based on data-driven choices of these thresholds.
    The results are based on 2,500 Monte Carlo iterations.}
    \label{fig:year_ahead_fixed}
\end{figure}

Figures~\ref{fig:quarter_ahead_fixed} and \ref{fig:year_ahead_fixed} report the performance of the `Old' and `New' methods for quarter-ahead and year-ahead predictions, both with and without the data-driven threshold choices. We display the results only for the sample size of $T = 500$, as the qualitative patterns are similar for other sample sizes. In each row of each figure, the left panel shows the results in which the `New' method uses fixed thresholds, whereas the right panel shows the results in which the `New' method uses the data-driven choice.


We observe that the results do not differ substantially between the `New' method based on fixed thresholds and the `New' method based on the data-driven choice. The data-driven choice performs slightly better for more extreme quantiles, which is natural given that these extreme quantiles lie farther away from the fixed thresholds set at the 90th and 10th percentiles.
The takeaway is that these threshold choices matter only marginally for moderately extreme quantiles, but their impact becomes more pronounced for truly extreme quantiles.
For robustness, we recommend using the data-driven choice.

\subsection{Constant Tail Exponents}\label{sec:additional_simulation:constant}

Recall that our motivation for proposing the `New' method stems from the limitations of the `Old' method discussed in Section~\ref{sec:limitations}. In particular, the `Old' method necessarily imposes a constant tail exponent $v(X_t)$ as a function of $X_t$. The simulation design used in Section~\ref{sec:designs} of the main text featured a tail exponent $v(X_t)$ that varies with $X_t$, thereby invalidating the `Old' method while validating the `New' one.

In this section, we present additional simulation results under a modified data-generating design in which the tail exponent $v(X_t)$ is constant in $X_t$, a setting that is more favorable to the `Old' method.

\subsubsection{Data Generating Designs}

Our data-generating designs remain essentially the same as those in Section~\ref{sec:designs} of the main text, except that the tail exponent model is modified as follows. 

For quarter-ahead predictions, recall that Section~\ref{sec:designs} uses the tail exponent model
$
v(X_t) = \exp(2.848 - 0.162 X_{t1} + 0.303 X_{t2}).
$
Instead, we use the constant tail exponent model
\begin{align*}
v(X_t) = \exp(2.848 - 0.162 \bar X_{t1} + 0.303 \bar X_{t2}),
\end{align*}
where $\bar X_{t1}$ and $\bar X_{t2}$ denote the sample means of $X_{t1}$ and $X_{t2}$ from the data on which the simulation design is calibrated.

Similarly, for year-ahead predictions, we replace the original model
$
v(X_t) = \exp(1.214 + 0.115 X_{t1} + 0.340 X_{t2})
$
with the constant tail exponent model
\begin{align*}
v(X_t) = \exp(1.214 + 0.115 \bar X_{t1} + 0.340 \bar X_{t2}).
\end{align*}

With these modifications, the tail exponent $v(X_t)$ becomes constant in $X_t$. This, in particular, implies that the limitations of the `Old' method discussed in Section~\ref{sec:limitations} of the main text do not apply to these specific data-generating designs.

\subsubsection{Simulation Results}

\begin{figure}[tbp]
    \centering
    \includegraphics[width=0.49\textwidth]{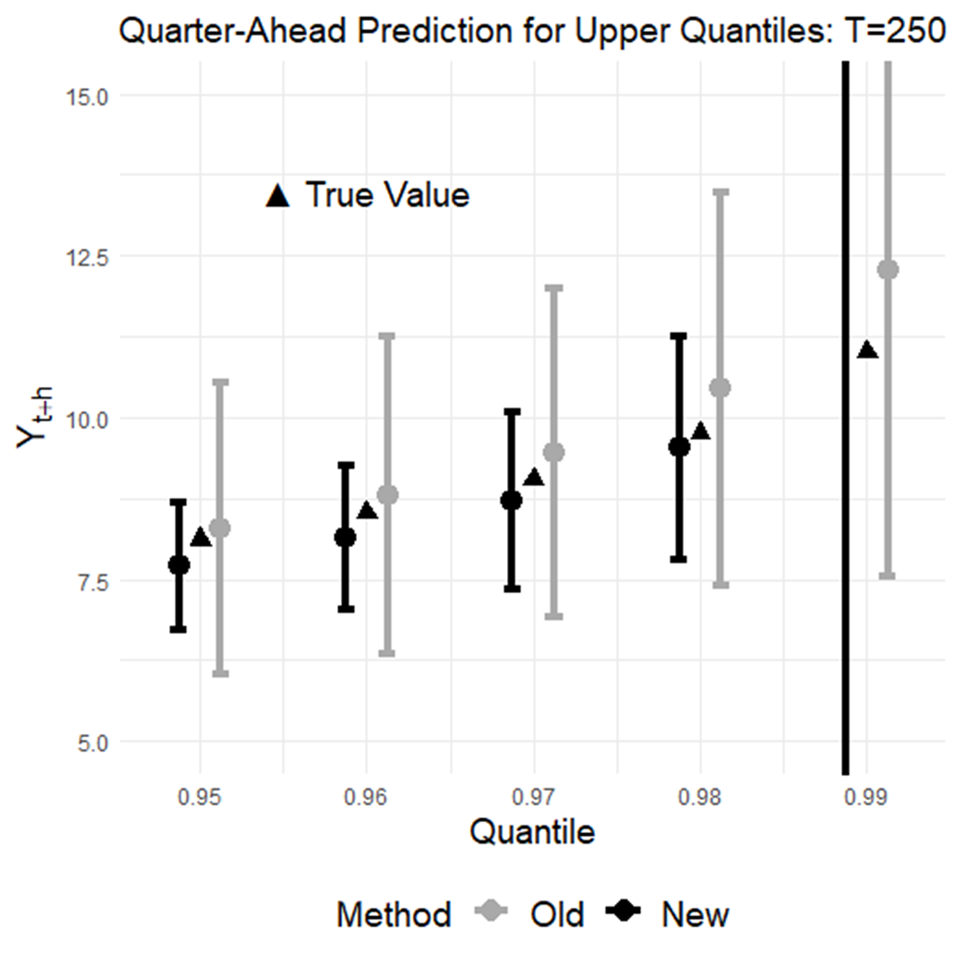}
    \includegraphics[width=0.49\textwidth]{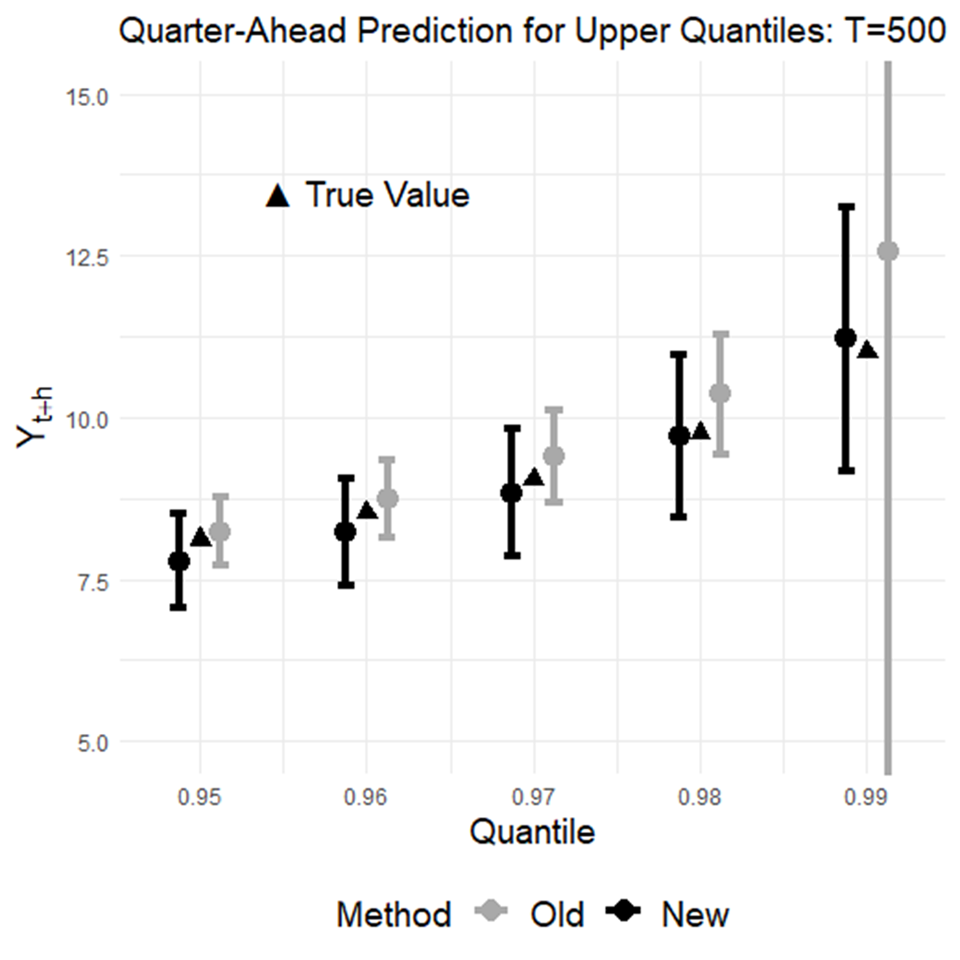}
    \\${}$\\
    \includegraphics[width=0.49\textwidth]{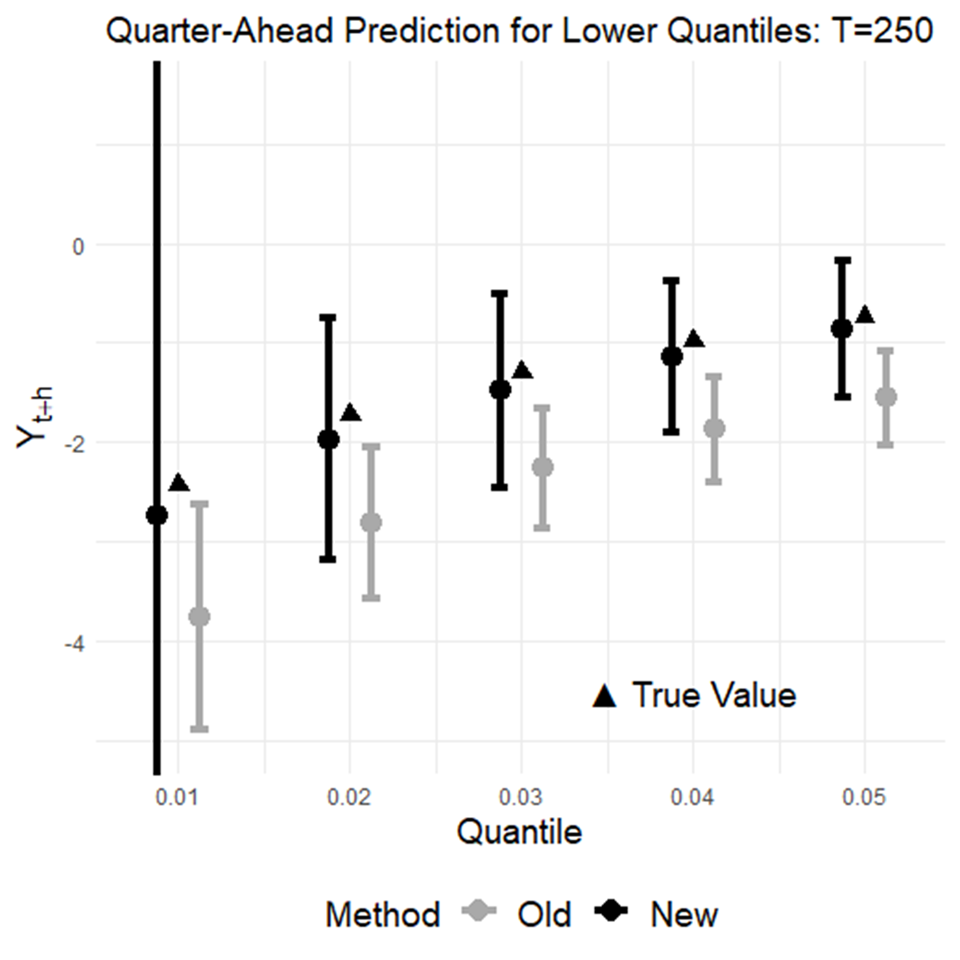}
    \includegraphics[width=0.49\textwidth]{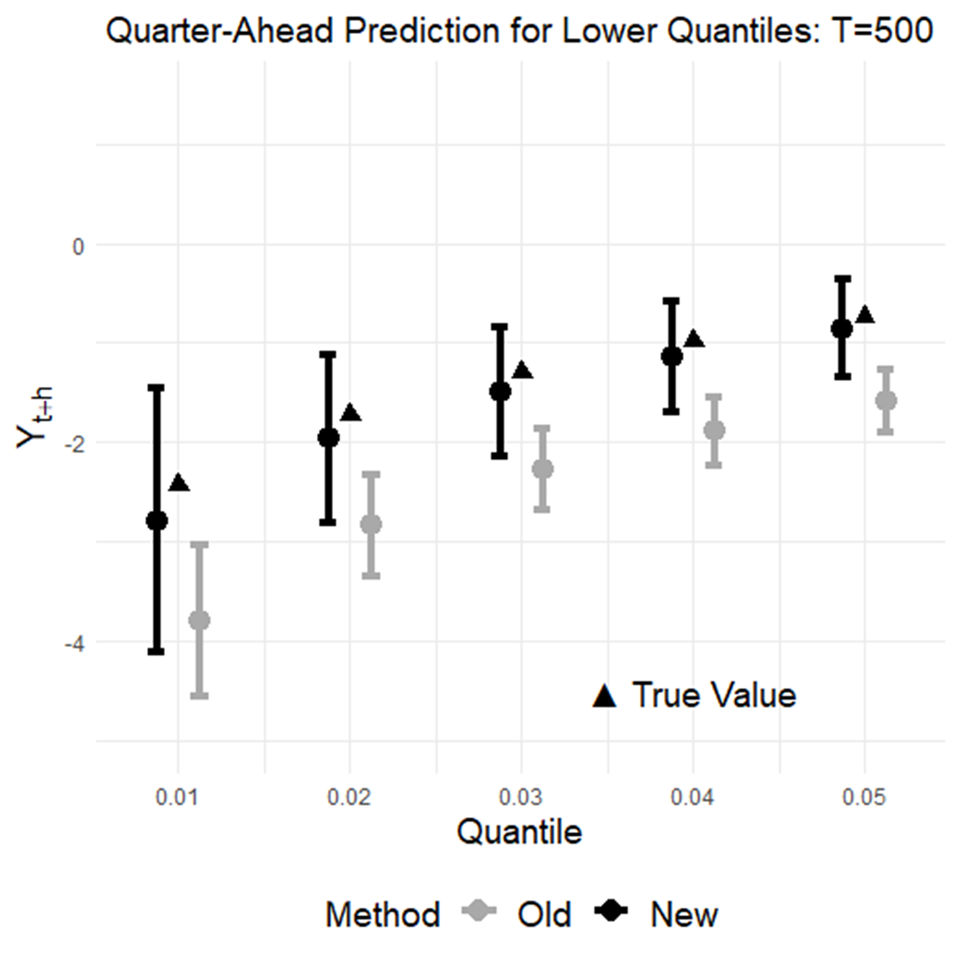}
    \caption{Simulation results comparing the performance of the proposed method (in black) and the existing method (in gray) for quarter-ahead predictions of growths. The predictions are conditional on the average values of $(X_{t1},X_{t2})= x_0 :=(2.732,0.007)$. The upper panels present results for the upper tail ($\tau \in \{0.95,0.96,0.97,098,0.99\}$), while the lower panels show results for the lower tail ($\tau \in \{0.01,0.02,0.03,0.04,0.05\}$). Dots represent simulation averages, bars represent the Gaussian interquartile ranges, and triangles denote the true values. The left column illustrates results for $T=250$, and the right column illustrates results for $T=500$. The results are based on 2,500 Monte Carlo iterations.}
    \label{fig:quarter_ahead_const}
\end{figure}

\begin{figure}[tbp]
    \centering
    \includegraphics[width=0.49\textwidth]{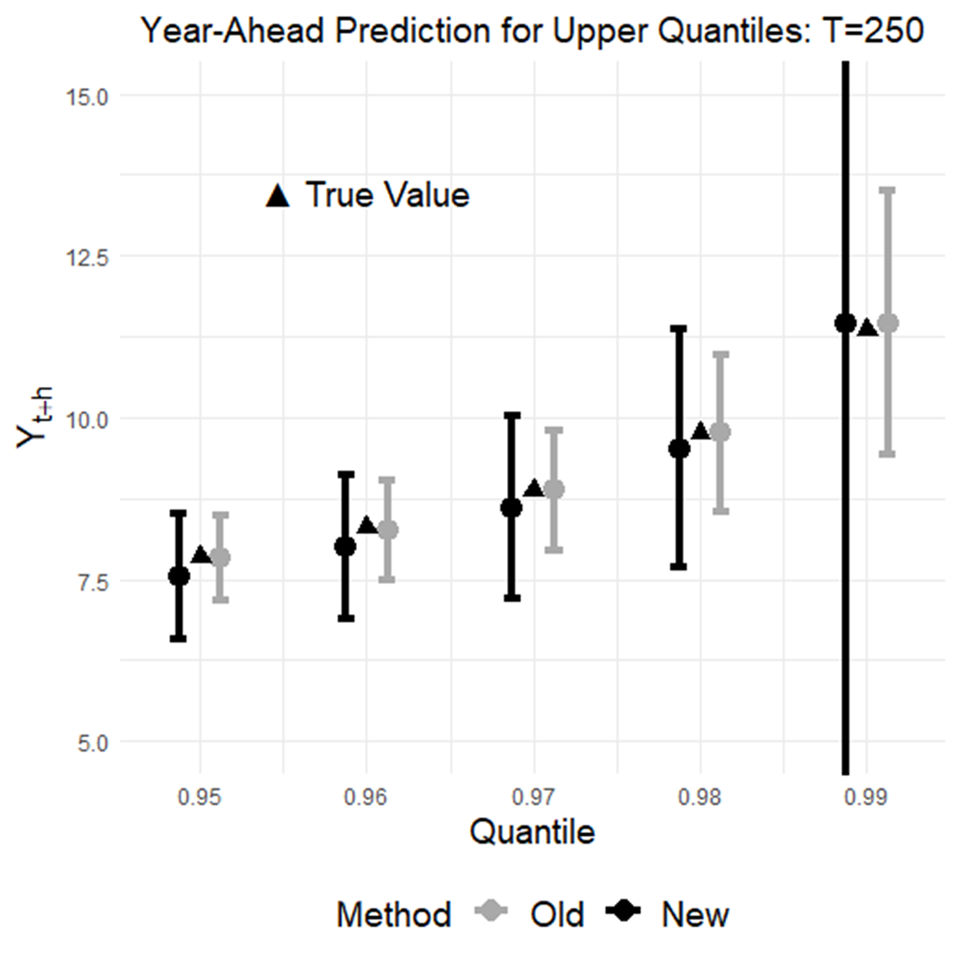}
    \includegraphics[width=0.49\textwidth]{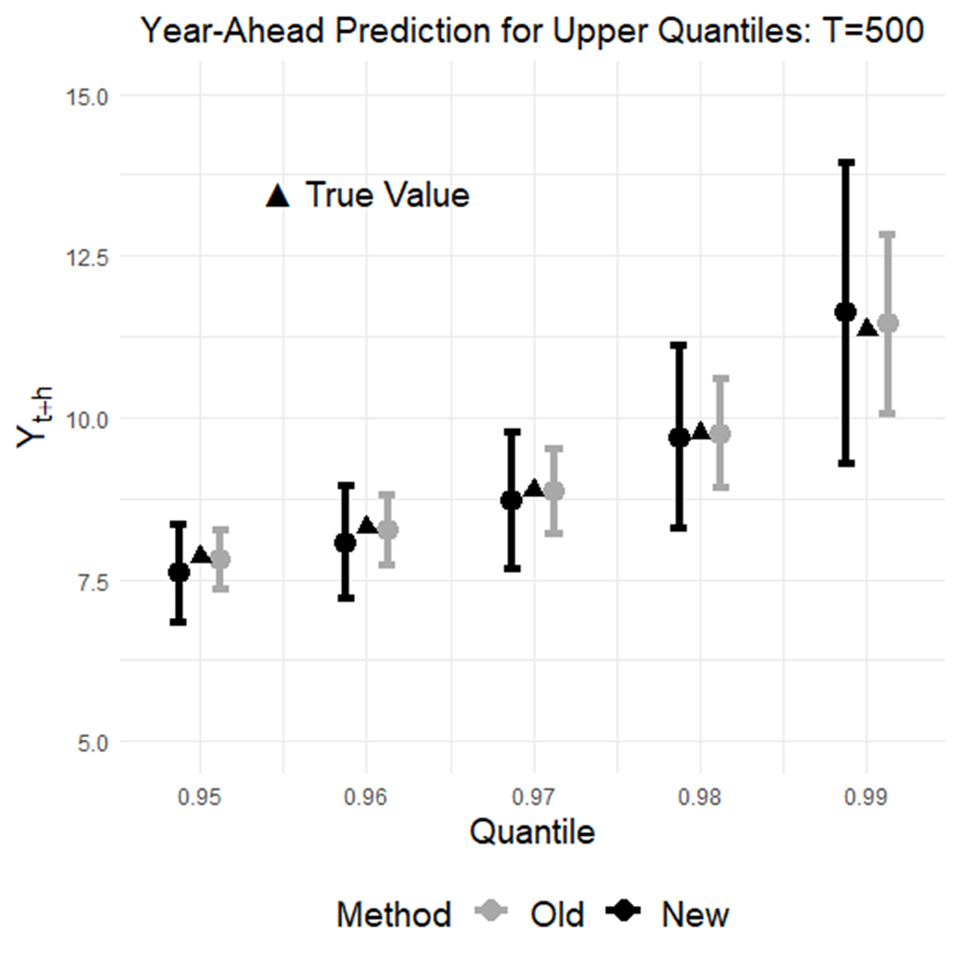}
    \\${}$\\
    \includegraphics[width=0.49\textwidth]{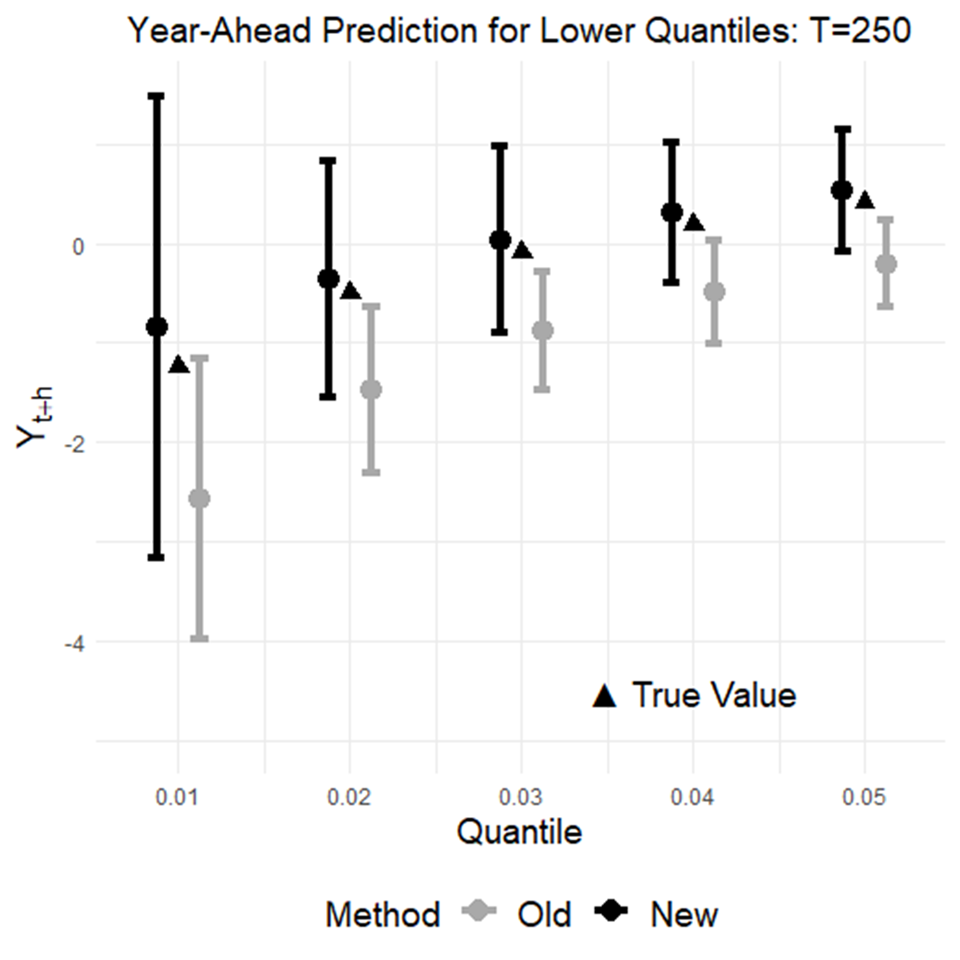}
    \includegraphics[width=0.49\textwidth]{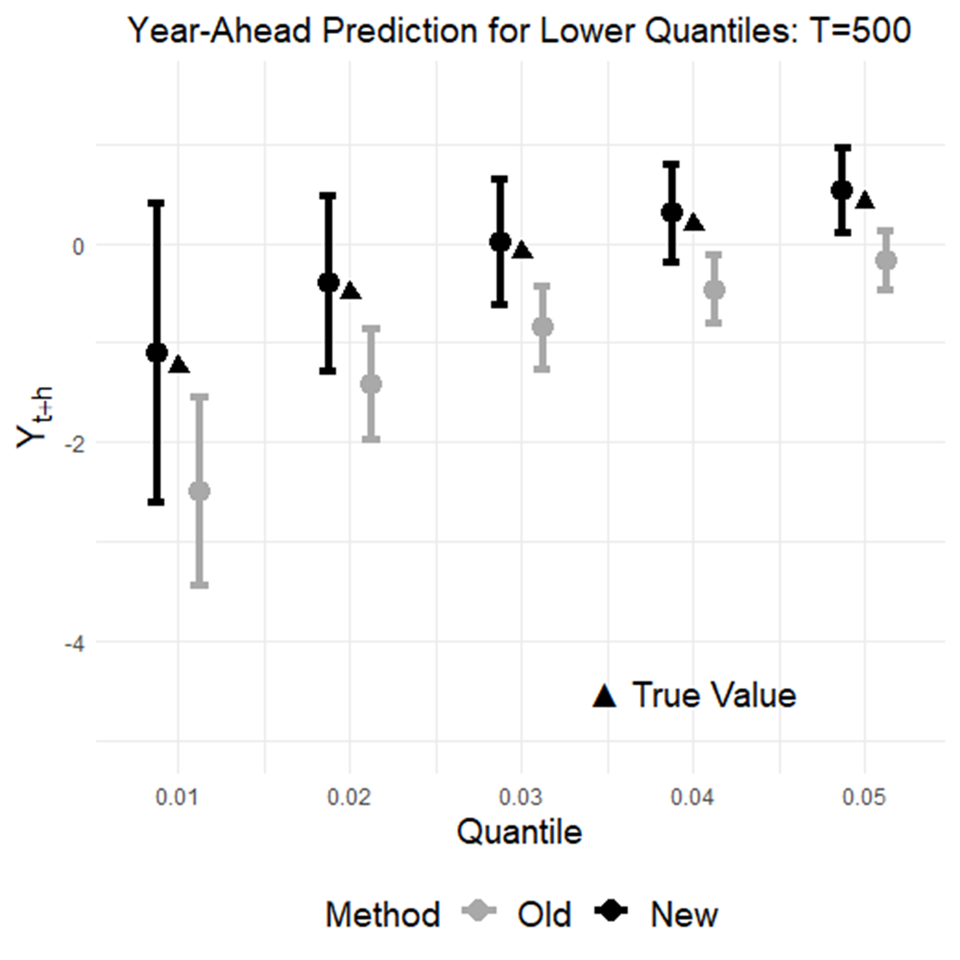}
    \caption{Simulation results comparing the performance of the proposed method (in black) and the existing method (in gray) for year-ahead predictions of growths. The predictions are conditional on the average values $(X_{t1},X_{t2})=x_0 :=(2.761,0.018)$. The upper panels present results for the upper tail ($\tau \in \{0.95,0.96,0.97,098,0.99\}$), while the lower panels show results for the lower tail ($\tau \in \{0.01,0.02,0.03,0.04,0.05\}$). Dots represent simulation averages, bars represent the Gaussian interquartile ranges, and triangles denote the true values. The left column illustrates results for $T=250$, and the right column illustrates results for $T=500$. The results are based on 2,500 Monte Carlo iterations.}
    \label{fig:year_ahead_const}
\end{figure}

Figures \ref{fig:quarter_ahead_const} and \ref{fig:year_ahead_const} illustrate the simulation results for quarter-ahead predictions ($h=1$) and year-ahead predictions ($h=4$), respectively.

Observe that Figures~\ref{fig:quarter_ahead_const} and \ref{fig:year_ahead_const} show that the `Old' method performs slightly better than in the baseline model with a non-constant tail exponent $v(X_t)$, as illustrated in Figures~\ref{fig:quarter_ahead} and \ref{fig:year_ahead} of the main text. This improvement arises because the constant tail exponent model considered here removes the issue associated with the limitations of the `Old' method, as discussed in Section~\ref{sec:limitations} of the main text.

Nonetheless, the `Old' method continues to underperform relative to our proposed `New' method even under this favorable setting. We therefore recommend using the more robust `New' method even in settings where the researcher knows that the tail exponent is constant across $X_t$.

\subsection{Misspecification}\label{sec:additional_simulation:misspecification}

The simulation design used in Section~\ref{sec:designs} of the main text features a tail exponent $v(X_t)$ that is linear in $X_t$, consistent with the modeling assumption of our `New' method. In this section, we present additional simulation results under a modified data-generating design in which the tail exponent $v(X_t)$ is nonlinear in $X_t$, a setting that leads to misspecification of both our `New' method and the `Old' method.

\subsubsection{Data Generating Designs}

Our data-generating designs remain essentially the same as those in Section~\ref{sec:designs} of the main text, except that the tail exponent model is modified as follows. 

For quarter-ahead predictions, recall that Section~\ref{sec:designs} uses the tail exponent model
$
v(X_t) = \exp(2.848 - 0.162 X_{t1} + 0.303 X_{t2}).
$
Instead, we use the nonlinear (-in-$X_t$) tail exponent model
\begin{align*}
v(X_t) = \exp(2.848 - 0.162 X_{t1} + 0.303 X_{t2} + c (X_{t1}^2 + X_{t2}^2)),
\end{align*}
where we set $c=0.1$.

Similarly, for year-ahead predictions, we replace the original model
$
v(X_t) = \exp(1.214 + 0.115 X_{t1} + 0.340 X_{t2})
$
with the nonlinear (-in-$X_t$) tail exponent model
\begin{align*}
v(X_t) = \exp(1.214 + 0.115 X_{t1} + 0.340 X_{t2} + c (X_{t1}^2 + X_{t2}^2)).
\end{align*}

With these modifications, the linear tail exponent model for $v(X_t)$ becomes misspecified for both the `Old' and `New' methods.

\subsubsection{Simulation Results}

\begin{figure}[tbp]
    \centering
    \includegraphics[width=0.49\textwidth]{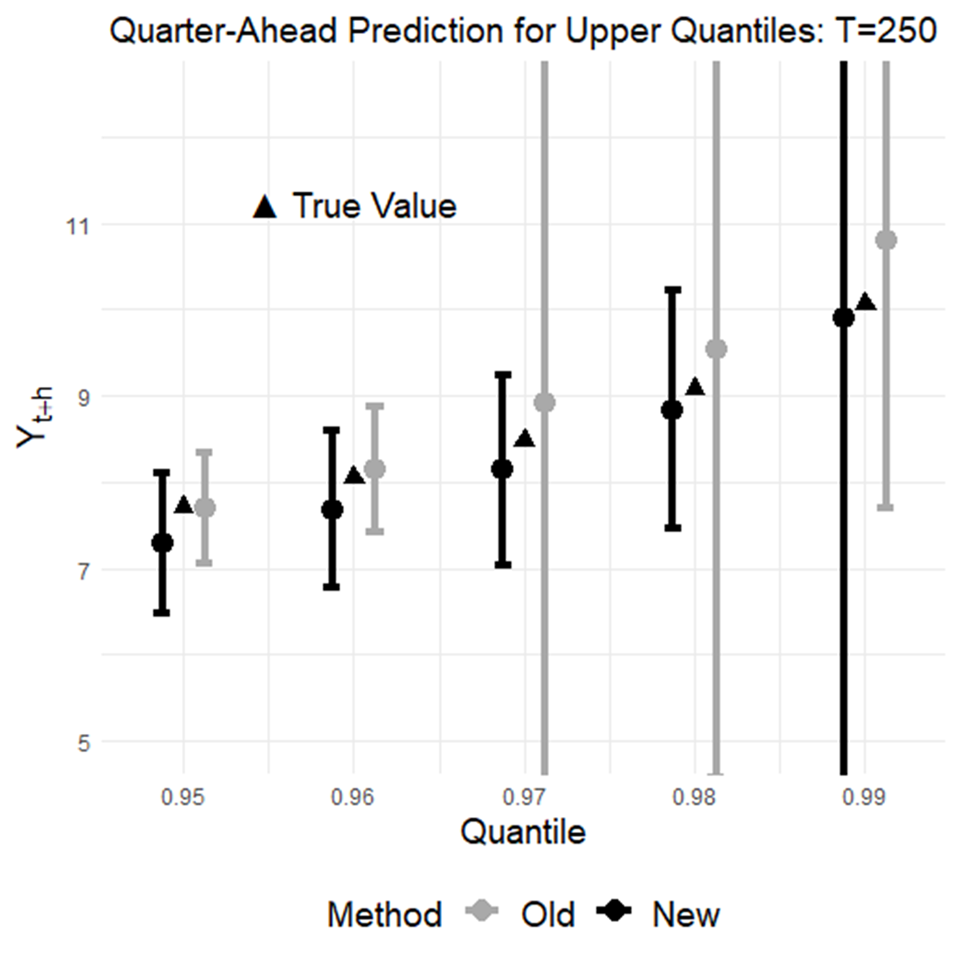}
    \includegraphics[width=0.49\textwidth]{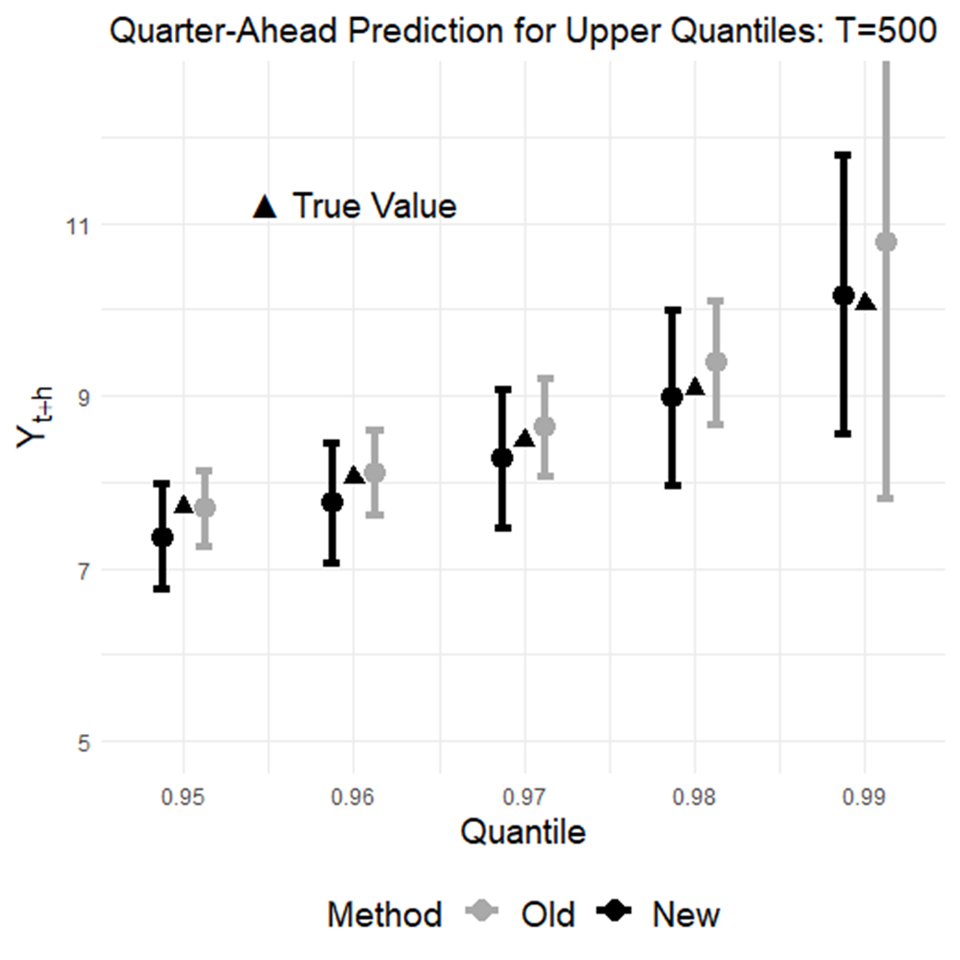}
    \\${}$\\
    \includegraphics[width=0.49\textwidth]{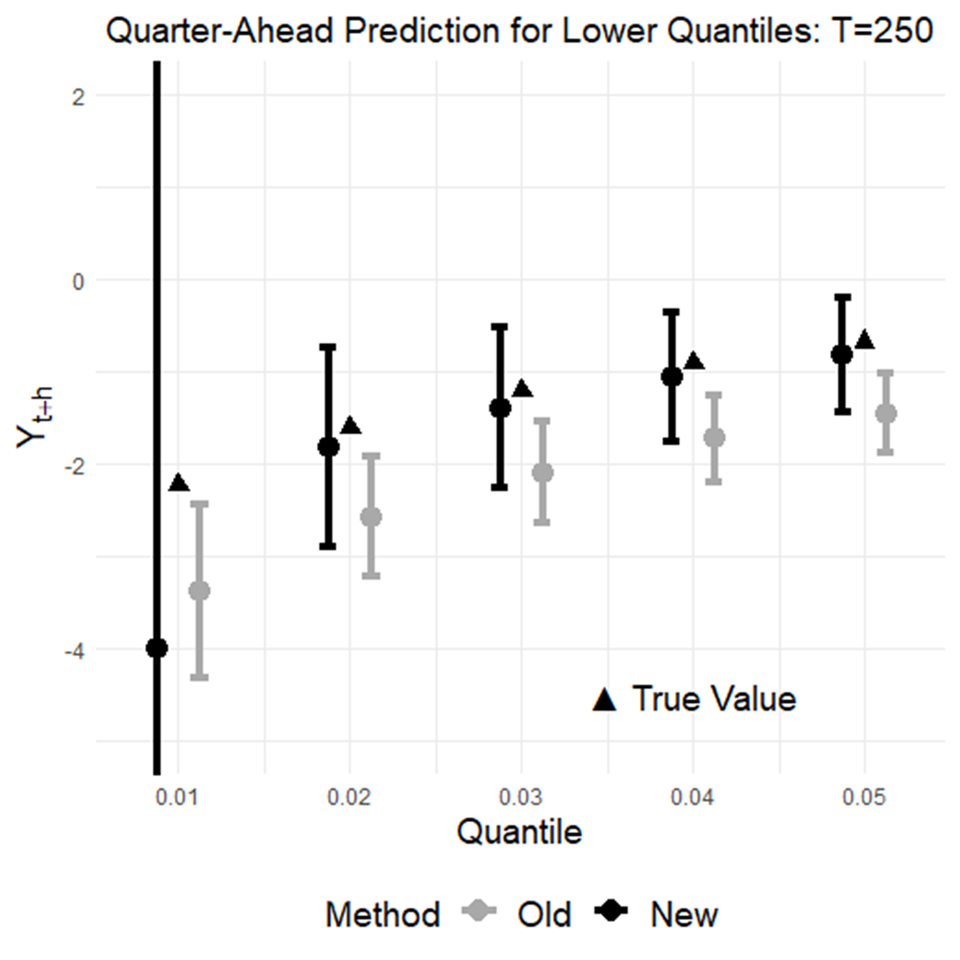}
    \includegraphics[width=0.49\textwidth]{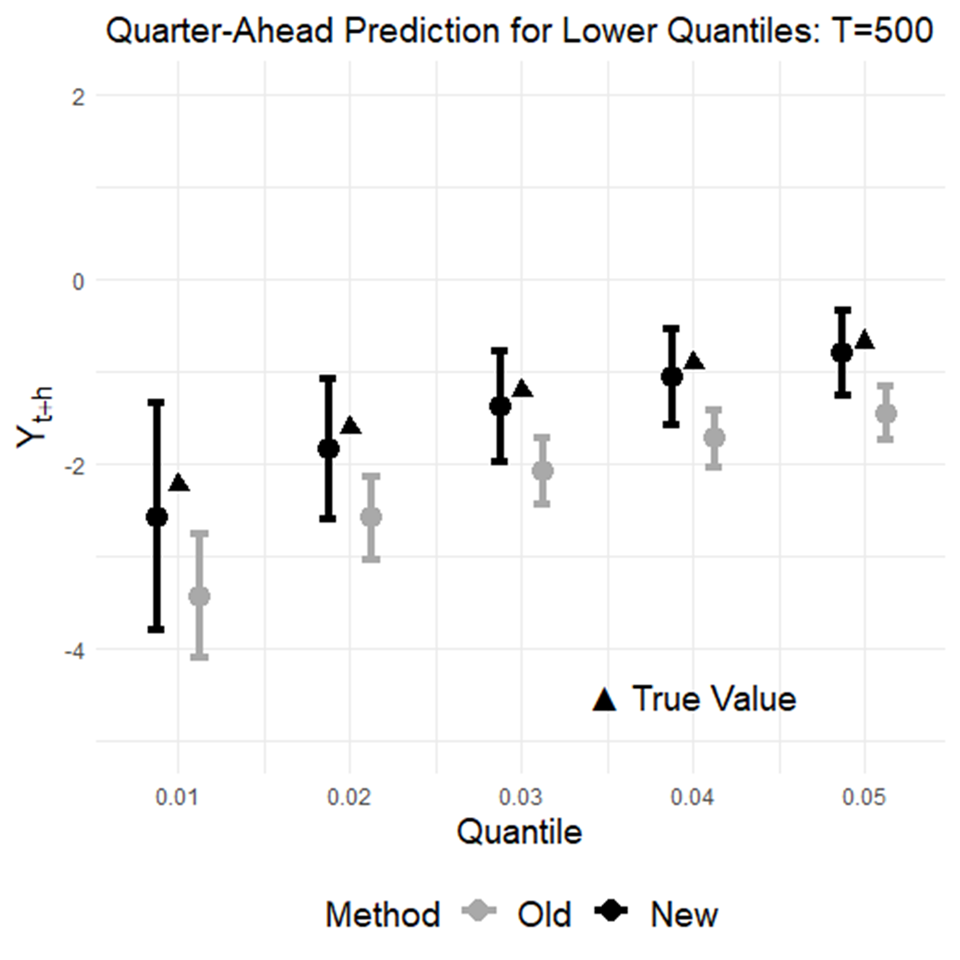}
    \caption{Simulation results comparing the performance of the proposed method (in black) and the existing method (in gray) for quarter-ahead predictions of growths. The predictions are conditional on the average values of $(X_{t1},X_{t2})= x_0 :=(2.732,0.007)$. The upper panels present results for the upper tail ($\tau \in \{0.95,0.96,0.97,098,0.99\}$), while the lower panels show results for the lower tail ($\tau \in \{0.01,0.02,0.03,0.04,0.05\}$). Dots represent simulation averages, bars represent the Gaussian interquartile ranges, and triangles denote the true values. The left column illustrates results for $T=250$, and the right column illustrates results for $T=500$. The results are based on 2,500 Monte Carlo iterations.}
    \label{fig:quarter_ahead_miss}
\end{figure}

\begin{figure}[tbp]
    \centering
    \includegraphics[width=0.49\textwidth]{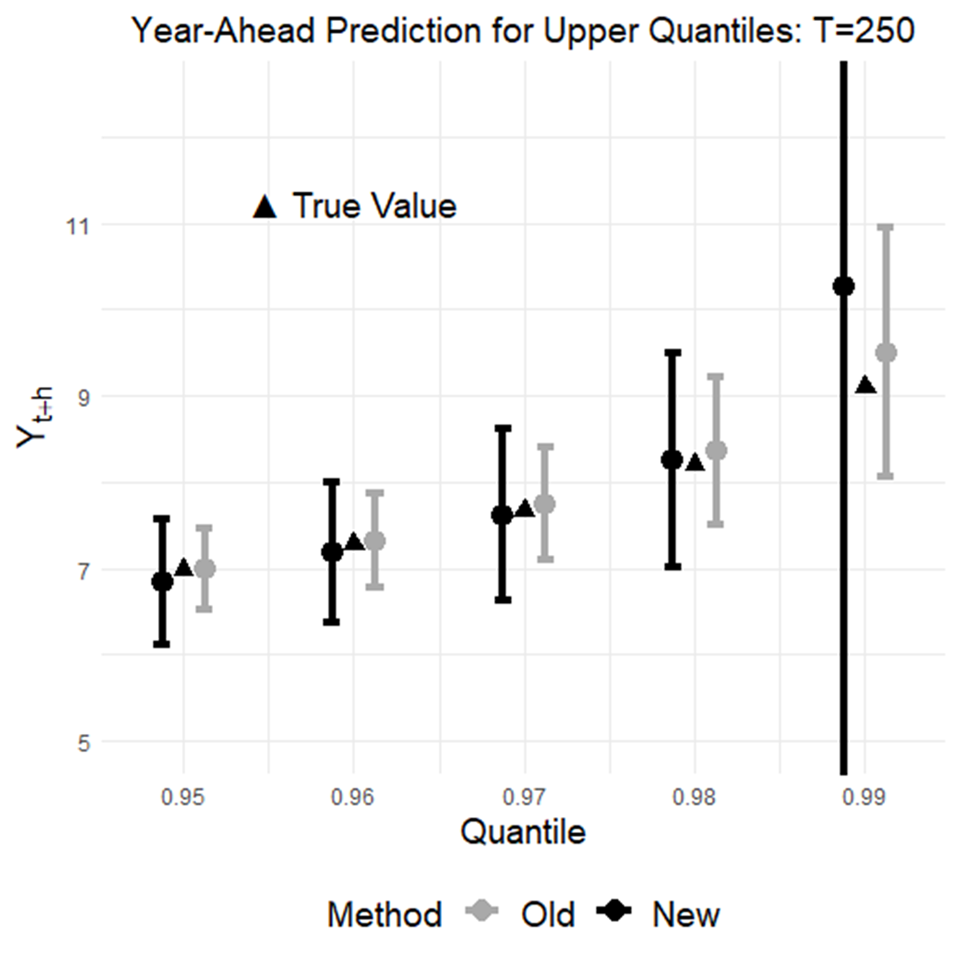}
    \includegraphics[width=0.49\textwidth]{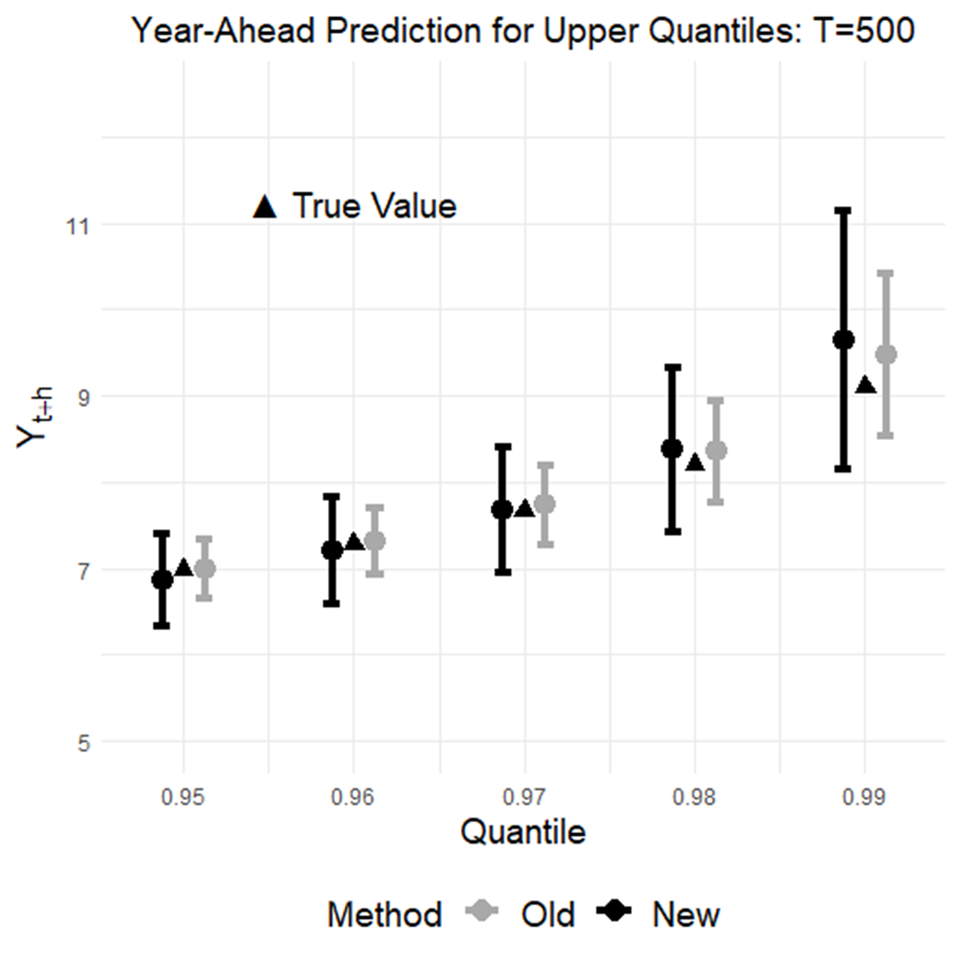}
    \\${}$\\
    \includegraphics[width=0.49\textwidth]{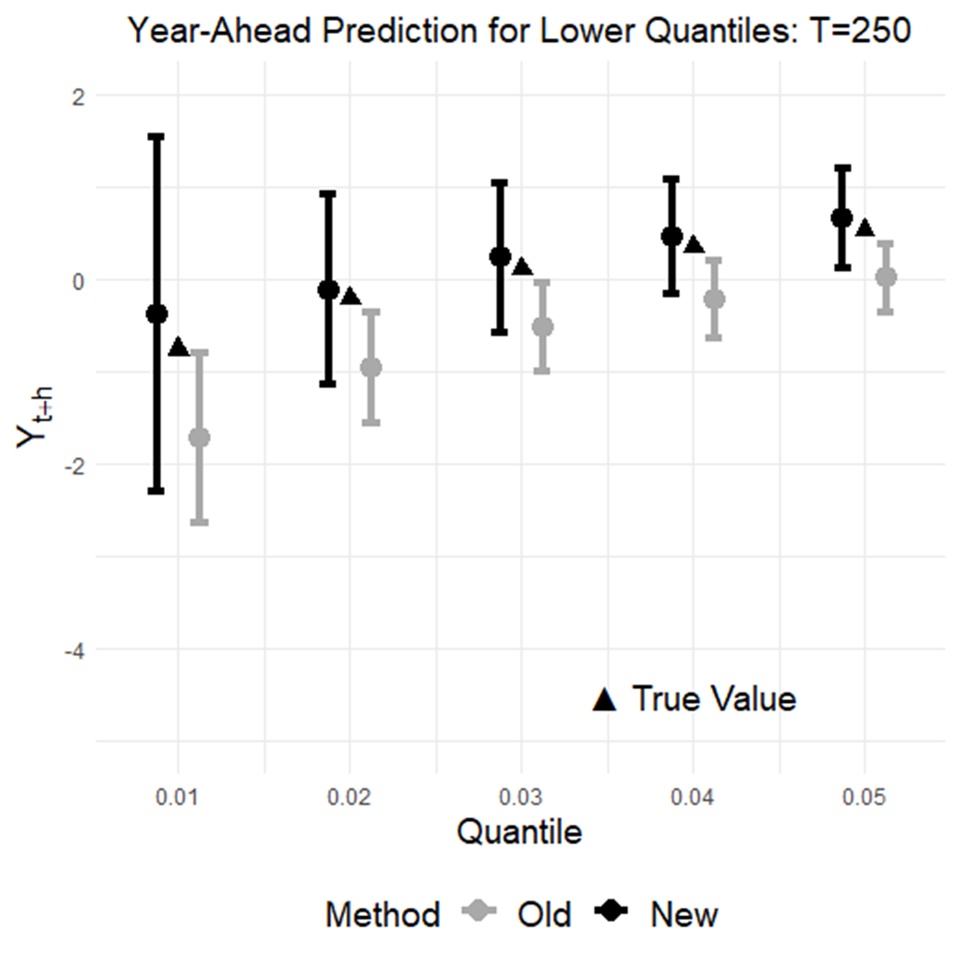}
    \includegraphics[width=0.49\textwidth]{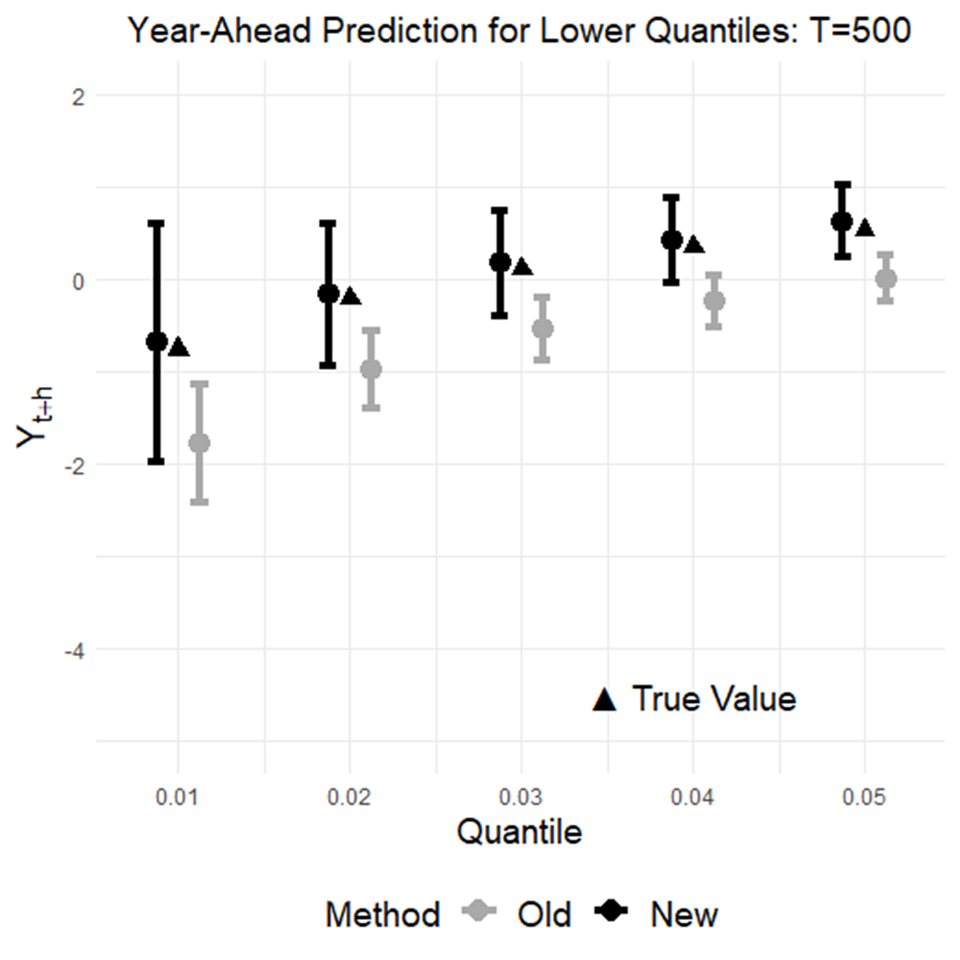}
    \caption{Simulation results comparing the performance of the proposed method (in black) and the existing method (in gray) for year-ahead predictions of growths. The predictions are conditional on the average values $(X_{t1},X_{t2})=x_0 :=(2.761,0.018)$. The upper panels present results for the upper tail ($\tau \in \{0.95,0.96,0.97,098,0.99\}$), while the lower panels show results for the lower tail ($\tau \in \{0.01,0.02,0.03,0.04,0.05\}$). Dots represent simulation averages, bars represent the Gaussian interquartile ranges, and triangles denote the true values. The left column illustrates results for $T=250$, and the right column illustrates results for $T=500$. The results are based on 2,500 Monte Carlo iterations.}
    \label{fig:year_ahead_miss}
\end{figure}

Figures \ref{fig:quarter_ahead_miss} and \ref{fig:year_ahead_miss} illustrate the simulation results for quarter-ahead predictions ($h=1$) and year-ahead predictions ($h=4$), respectively.

Compared to the simulation results illustrated in Figures~\ref{fig:quarter_ahead} and \ref{fig:year_ahead} for the baseline model with correct specification of the tail exponent, the precision of the predictions, both in terms of biases and the lengths of the interquartile ranges,differs in Figures~\ref{fig:quarter_ahead_miss} and \ref{fig:year_ahead_miss} for both the `New' and `Old' methods. With this said, the qualitative behaviors remain largely the same as in the baseline simulation results, and the same interpretations continue to apply. In particular, the `New' method still overall outperforms the `Old' method. We experimented with other forms of misspecification as well, and the qualitative outcomes remained unchanged.

\subsection{Summary of the Additional Simulation Studies}

This appendix section presented three sets of additional simulation studies. 

First, Appendix~\ref{sec:additional_simulation:fixed} reports results obtained when constant threshold parameters $y_{\min}$ and $y_{\max}$ are used in place of the data-driven rule. While the results are largely similar, the data-driven choice performs better, especially for more extreme quantiles. These findings suggest that the data-driven threshold selection rule is preferable for robustness in general.

Second, Appendix~\ref{sec:additional_simulation:constant} explores simulation designs in which the tail exponent model $v(X_t)$ is defined to be constant in $X_t$. Even though this setting is favorable to the `Old' method, the `New' method continues to outperform it.

Third, Appendix~\ref{sec:additional_simulation:misspecification} considers misspecified tail exponent models $m(X_t)$, under which the model underlying the `New' method is itself misspecified. Despite this misspecification, the `New' method again continues to outperform the `Old' method.

Taken together, the latter two sets of simulation exercises highlight the robustness properties of the `New' method proposed in this paper.

\color{black}
\bibliographystyle{ecta}
\bibliography{mybib}
\end{document}